# A method to quantify molecular diffusion within thin solvated polymer films: A case study on films of natively unfolded nucleoporins

Rickard Frost[1], Delphine Débarre[2], Saikat Jana[1], Fouzia Bano[1], Jürgen Schünemann[3], Dirk Görlich[3] and Ralf P. Richter[1,*]

[1] School of Biomedical Sciences, Faculty of Biological Sciences, School of Physics and Astronomy, Faculty of Engineering and Physical Sciences, Astbury Centre of Structural Molecular Biology, and Bragg Centre for Materials Research, University of Leeds, Leeds, LS2 9JT, United Kingdom

[2] Univ. Grenoble Alpes, CNRS, LIPhy, 38000 Grenoble, France

[3] Department of Cellular Logistics, Max Planck Institute for Biophysical Chemistry, 37077 Göttingen, Germany

* Corresponding author: r.richter@leeds.ac.uk

**Abstract**

We present a method to probe molecular and nanoparticle diffusion within thin, solvated polymer coatings. The device exploits the confinement with well-defined geometry that forms at the interface between a planar and a hemi-spherical surface (of which at least one is coated with polymers) in close contact, and uses this confinement to analyse diffusion processes without interference of exchange with and diffusion in the bulk solution. With this method, which we call plane-sphere confinement microscopy (PSCM), information regarding the partitioning of molecules between the polymer coating and the bulk liquid is also obtained. Thanks to the shape of the confined geometry, diffusion and partitioning can be mapped as a function of compression and concentration of the coating in a single experiment. The method is versatile and can be integrated with conventional optical microscopes, and thus should find widespread use in the many application areas exploiting functional polymer coatings. We demonstrate the use of PSCM using brushes of natively unfolded nucleoporin domains rich in phenylalanine−glycine repeats (FG domains). A meshwork of FG domains is known to be responsible for the selective transport of nuclear transport receptors (NTR) and their macromolecular cargos across the nuclear envelope that separates the cytosol and the nucleus of living cells. We find that the selectivity of NTR uptake by FG domain films depends sensitively on FG domain concentration, and that the interaction of NTRs with FG domains obstructs NTR movement only moderately. These observations contribute important information to better understand the mechanisms of selective NTR transport.

Solvated polymer films at the solid-liquid interface constitute a wide span of surface coatings, including polymers physically adsorbed or grafted to/from a supporting surface (planar, structured, or particulate). Such polymer films may be either passive or responsive to external stimuli, *e.g.* changes in temperature, pH, ionic strength or light. To physically adsorb polymers onto a solid surface is a simple surface functionalization procedure and can be accomplished using methods like dip-coating etc. Similarly, more advanced surface adlayers may be built layer-by-layer through sequential exposure to oppositely charged polyions. Grafting of polymers to/from a solid support requires more specific surface chemistry approaches but generally results in a more durable surface coating. An important sub-category of surface grafted polymers are polymer brushes. In such surface coatings, polymers are one-end grafted at high density to the solid support, forming a brush like structure.[1,2] Independent of the surface functionalization strategy, confinement of polymers in a surface associated layer affects their conformation and self-organisation. Thus, the properties of the polymer are different when associated to a surface compared to when present in bulk solution. Furthermore, and importantly, the polymer film may significantly alter and enable tuning of the properties of the solid surface.

During the last few decades, solvated polymer coatings have been investigated for a broad range of applications, from fundamental research to everyday-life applications. Examples include reconstituted biomolecular and biomimetic films,[3] biomaterials,[4] biosensors,[5] nanomedicine, anti-fouling and anti-microbial coatings,[6,7] purification and separation membranes, food processing, paints, lubrication[8], and energy storage.[9] An important functional parameter of such coatings is how the constituent polymers and active substances (*e.g.* active synthetic molecules, proteins, nanoparticles, viruses; here collectively called solutes) diffuse within them. Depending on the application, one may design ways to either enhance or delay such diffusion. Consequently, there is a broad need for the analysis and quantification of diffusion processes within thin polymer films.

However, this is currently challenging when the film is immersed in a solvent phase. Methods based on optical microscopy, such as fluorescence recovery after photobleaching (FRAP), fluorescence correlation spectroscopy (FCS) and single particle tracking (SPT),[10] are well-established and popular to study diffusion processes. FRAP and FCS though fail for thin films that dynamically exchange solutes with the bulk solution: because the dimensions of the volume probed (> 200 nm in *xy*, and > 500 nm in *z*, for diffraction-limited confocal optics) exceed the film thickness (≲ 100 nm), diffusion scenarios including (i) diffusion within the film, (ii) diffusion in the adjacent bulk liquid, and (iii) exchange between the film and the bulk all contribute to the optical signal (as illustrated in Figure 1A); in this scenario, it is thus challenging to separate in-film diffusion from the other two processes. Similarly, although single particle tracking is able to determine diffusion coefficients in smaller spaces, statistical analysis becomes limited when diffusion trajectories within the film are short owing to rapid exchange between the film and the bulk solution.

Here, we present an analytical methodology, based on optical microscopy, that overcomes this limitation by confining polymer film(s) between two surfaces, one planar and the other macroscopically curved (as illustrated in Figure 1B). The confined volume near the contact point retains nanometre dimensions along the optical axis, inferior or comparable to the thickness of the polymer film yet at the same time the lateral dimensions are micrometric and thus large enough to be resolved with conventional microscopy. The setup effectively excludes the bulk solution from a region close to the surface-surface contact so that solute diffusion within the film can be probed and confounding solute exchanges with the bulk are excluded. This concept solves an important problem in thin film analysis for which there is currently no solution. Using the same approach, partitioning of solutes between the polymer film and the bulk liquid can also be readily measured. In addition, because the contact force between the surfaces can be set and the geometry of the interface is known, a defined gradient of polymer compression and concentration is created and it becomes possible to measure solute diffusion and partitioning as a function of these parameters in a single measurement. This substantially extends the capability of the methodology.

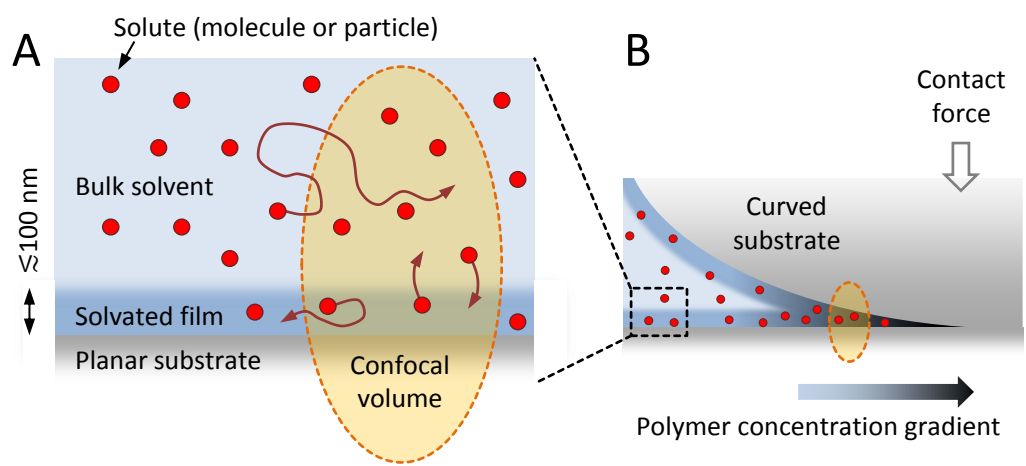

**Figure 1. Schematic representation of the problem at hand and the experimental approach. (A)** Probe solutes (red dots) partition between the bulk solution and a thin, solvated polymer film. They diffuse in these phases with rates $D_{bulk}$ and $D_{film}$, respectively. The objective is to quantify $D_{film}$. Because the probe molecules continuously move between the two phases, and the polymer film thickness is below the optical resolution limit, it is challenging to separate $D_{film}$ from $D_{bulk}$. **(B)** By confining the polymer coating(s) between a planar and a curved surface, the bulk solution is excluded in the region surrounding the contact between the two surfaces. This enables optical microscopy to probe diffusion within the polymer film(s), and also to quantify the partitioning of probe molecules between the bulk and polymer phases. The gradual compression of the polymer film(s) near the contact point also entails a polymer concentration gradient that can be exploited to measure the diffusion and partitioning of probe molecules as a function of polymer film compression and concentration. Note that the contact geometry was stretched along the vertical axis for illustrative purposes: in reality, the gap height increases very slowly with distance from the contact point, and the confocal volume will always include the entire thickness of the gap across the imaged area.

It should be noted here that the generation of confined spaces using a plane-sphere geometry is a well-known procedure. The surface force apparatus, for example, combines this geometry with exquisite sensitivity in force and separation distance,[11] to measure interaction forces between functionalised surfaces; more recently, this approach has also been combined with optical analysis for concurrent studies of the molecular organisation and diffusion within the confined fluids by microscopy and spectroscopy techniques.[12-15] Plane-sphere geometries have also already been combined with optical microscopy for improved single molecule imaging,[16-18] or to visualise dynamic processes under confinement as diverse as blood clot formation,[19] lubricant transfer during interfacial shear[20] and capillary condensation.[21] Distinct aspects of the here-described method are the application to solvated polymer films, and its ease of integration with existing microscopes and imaging modalities, where a substantially static and constant contact force is beneficial to control the compression of the polymer film.

The new methodology, which we call plane-sphere confinement microscopy (PSCM), uses reflection interference contrast microscopy (RICM), to analyse the gap profile between the apposed surfaces (which is defined by the shape of the surfaces, the applied load, and the thickness and compressibility of the polymer films); fluorescence microscopy, to image the distribution of probe molecules around the contact point; and FRAP to probe diffusion of probe molecules within the confined polymer films. We demonstrate the use of PSCM using model systems of the nuclear pore permselectivity barrier, an important biological confined polymer matrix which makes the transport of macromolecules between the cell nucleus and the cytoplasm highly selective.[22]

**Case study: the nuclear pore permselectivity barrier**

The presented analytical methodology, PSCM, is generic and applicable to a wide range of polymer film systems. To demonstrate the use of the methodology, we have selected a biomimetic system of the nuclear pore permselectivity barrier. Nuclear pore complexes (NPCs) control the exchange of biomolecules between the nucleus and the cytoplasm of all eukaryotic cells.[22] NPCs perforate the nuclear envelope, and through selective transport of RNA and proteins enable the spatial separation of transcription (cell nucleus) and translation (cytoplasm), which provides a powerful mechanism to control gene expression. Although small molecules up to roughly 5 nm in diameter can diffuse freely across the NPC, the passage of larger macromolecules is impeded unless they are bound to nuclear transport receptors (NTRs).[23] The NPC consists of a scaffold of folded proteins that defines an approximately 40 nm wide channel. The channel, however, is not empty but filled with a meshwork of specialized natively unfolded protein domains that are rich in phenylalanine-glycine (FG) dipeptides (FG domains), and acts as a selective permeation barrier.[24] For example, it has been shown that NTRs are substantially enriched in FG domain protein films.[25] NTRs tend to have many binding sites for FG dipeptide motifs, *i.e.* the interactions between NTRs and FG domains are intrinsically multivalent. Recent studies have found that the thermodynamic and morphological aspects of NTR binding to FG domain assemblies can be described well by simple models that consider the FG domains as homogeneous flexible polymers and the NTRs as featureless spheres. This indicates that detailed structural features of FG domains and NTRs are secondary to function and that simple soft matter physics models are able to capture essential features of the system.[25,26] Each FG domain contains several FG motifs, which contribute to intra- and intermolecular interactions of FG domains, as well as to the binding of NTRs. The attractive interactions between FG domains are essential for the functionality of the permeability barrier.[27] Permselectivity consists of three basic, sequential steps: (I) entry into the pore, (II) diffusion through the pore, and (III) release from the pore. While steps (I) and (III) have been studied in detail and begun to be understood, much less is known about step (II). Studies using intact nuclear pores have shown that translocation can occur fast (within milliliseconds).[28-30] Detailed analysis of single molecule tracks by Yang and Musser[29] revealed diffusion of a selected NTR-cargo when interacting with the nuclear pore complex is only moderately (*i.e.* less than tenfold) reduced compared to diffusion in the cytoplasm. Although it has been revealed that *individual* NTR-FG motif interactions are extremely fast,[31] it still remains to be determined how diffusion through the channel can occur rapidly with respect to *collective* NTR-FG motif interactions (*i.e.* the multivalent interactions between a given NTR and the FG motifs presented by the surrounding meshwork of FG domains). A main impediment within this area of research has been that techniques are lacking to study the diffusion process within confined spaces such as the NPC or other nanoscale phases.

*A reconstituted model of the nuclear pore permselectivity barrier.* Films of end-grafted FG domains (such as FG domain brushes)[3,32,33] have been successfully used as a model system to study the properties and mechanisms of function of the nuclear pore permselectivity barrier. Past work using this model system mainly focused on morphology[3,32,34] (*e.g.* film thickness changes and phase formation) and thermodynamic parameters[3,25,35,36] (*e.g.* partitioning of NTRs between the bulk phase and the FG domain film). Here, we use PSCM to extract information regarding both partitioning and diffusion of probe molecules within FG domain films. As probe molecules we utilize enhanced GFP (GFP$^{Std}$), a GFP mutant designed not to bind to FG domains (GFP$^{Inert}$), and a GFP mutant designed to gain NTR-like properties (GFP$^{NTR}$).[37] We thus demonstrate the use of PSCM and, for the first time to our knowledge, quantify the diffusion coefficient of an NTR-like protein within an ultrathin film of FG domains.

## RESULTS

We introduce plane-sphere confinement microscopy (PSCM) with the purpose to allow studies of diffusion processes within solvated polymer films at the solid-liquid interface. A planar and a semi-spherical surface, both functionalized with the polymer film of interest, were brought into contact in a well-controlled fashion using a micromanipulator (Figure 2A). Thus, close to the point of contact between the planar and spherical surface the polymer films will overlap, excluding all bulk liquid. This is the region of primary interest for PSCM: thanks to the large size of the hemi-sphere, its lateral dimensions will exceed 10 µm for polymer coatings of > 10 nm in thickness (*vide infra*). Processes on this length scale can be readily resolved by fluorescence microscopy thus enabling the characterization of diffusion processes inside the polymer film without interference from the bulk solution. We will first demonstrate how the confined geometry is realized and characterized, and then describe how information regarding the diffusion coefficient of the fluorescent probe molecule (solute) and its partitioning between the polymer film and the bulk solution can be extracted.

**Defining the confined space between polymer-coated plane and sphere**

*Polymer coatings.* To anchor FG domains of Nsp1 ($FG^{Nsp1}$) to desired surfaces we exploited the specific binding of poly-histidine tags (located at the C-terminus of $FG^{Nsp1}$) to $Ni^{2+}$-EDTA moieties on the two surfaces. The process to prepare films of C-terminally grafted $FG^{Nsp1}$ in this way has previously been established[33] and was here validated by quartz crystal microbalance (QCM-D) on an identically functionalized reference sensor surface (Supporting Figure S1). In previous work, we also demonstrated how the surface density of FG domain films can be quantified by spectroscopic ellipsometry (SE), and that QCM-D and SE data can be correlated to estimate surface densities from the QCM-D response.[3,33,25] Building on this prior work, we estimate that the $FG^{Nsp1}$ film used here has a surface density of 5 ± 1 pmol/cm$^2$ (equivalent to a root-mean-square anchor distance of 5.9 ± 0.6 nm; see Supporting Figure S1 for details). Moreover, extensive analysis by atomic force microscopy nanoindentation, QCM-D and SE had previously revealed the thickness of films of C-terminally anchored $FG^{Nsp1}$ at around 5 pmol/cm$^2$ to be $d_{FG} \approx 30$ nm.[34] Thus the uncompressed FG domain film has a mass concentration of 107 mg/mL, and harbours a total molar concentration of 55 mM FG dipeptides (each Nsp1 FG domain features 33 FG dipeptides[25]).

The method of FG domain surface grafting was then transferred to planar glass cover slips and rods with a hemi-spherical cap. Both types of surfaces were functionalized *in situ* and kept in working buffer at all times during and after FG domain film formation. Aided by a micromanipulator, the rod with hemi-spherical cap was aligned with the microscope objective and then lowered towards the planar surface until contact was reached (Figure 2A). The alignment procedure allowed the contact point between the two surfaces to be positioned in the centre of the field of view upon first contact (Supporting Figure S2).

To confirm successful FG domain film formation in the PSCM setup, we incorporated 1 mol-% of $FG^{Nsp1}$ labelled with Atto488 at the free N-terminus, and visualized the surface coatings in plane-sphere confinement geometry using confocal microscopy (Figure 2B). The fluorescence micrograph did not show any appreciable features (except in and close to the contact area, *vide infra*) as expected for homogeneous FG domain films. To probe how the FG domain films on the two apposed surfaces compare, a circular area close to the contact point was first photo-bleached, the rod with hemi-spherical cap was then withdrawn, translated to the right by approximately 50 µm and brought back into contact (Figure 2C-D). This procedure revealed that $FG^{Nsp1}$ was present on both surfaces at comparable surface density because the bleaching effect was split in two equal parts where the total loss of intensity in the two spots (2 × 45%; inset Figure 2D) was identical to the total loss of intensity in the original spot (90%; inset Figure 2C).

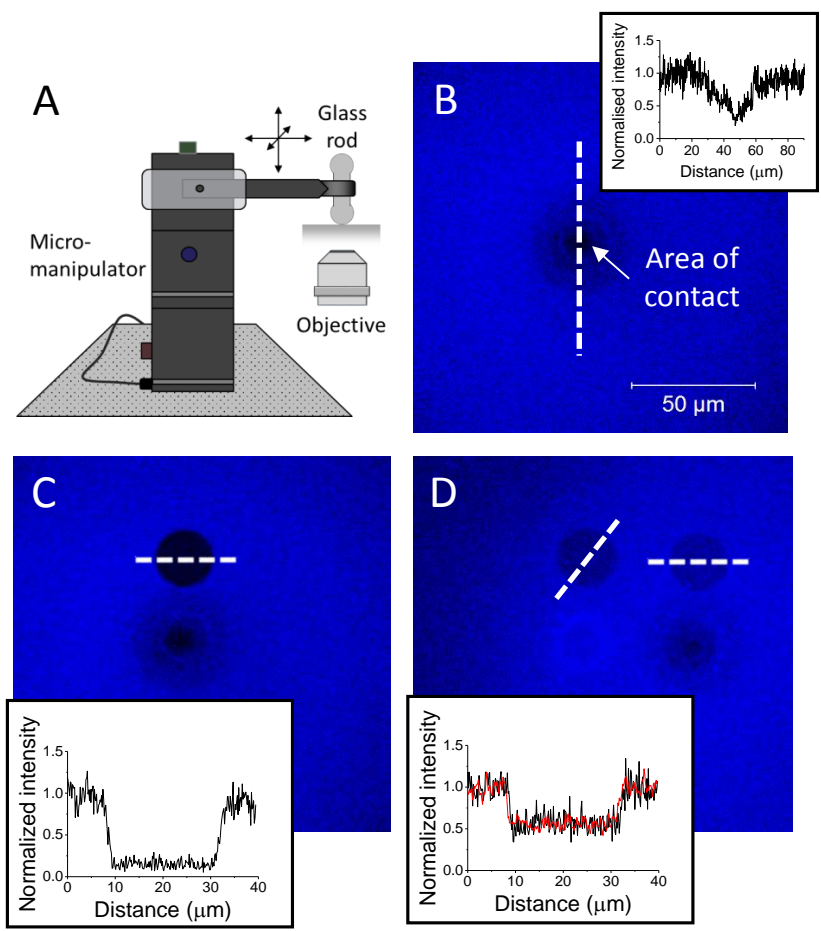

**Figure 2. Confinement of polymer films at the plane-sphere interface.** **(A)** Schematic representation of the experimental setup (not to scale). A glass rod with a hemi-spherical cap is coarse-aligned with the optical axis, and lowered towards a planar surface using a micromanipulator until contact is reached. **(B)** Fluorescence micrograph of the plane-sphere interface with both surfaces functionalized with $FG^{Nsp1}$-$His_{10}$ films (1 mol% of $FG^{Nsp1}$-$His_{10}$ was labelled with the fluorophore Atto488). The area of contact is visible as a zone of reduced fluorescence. Inset shows fluorescence intensity profile taken along the white dashed line. **(C)** Interface shown in (B) after photobleaching of a circular region. The lack of apparent recovery demonstrates that $FG^{Nsp1}$ was bound and immobile on the surfaces. **(D)** Interface shown in (C) after retracting the spherical surface and making a new contact approximately 50 μm to the right. Comparison of fluorescence intensity profiles (insets in (C) and (D), taken along the white dashed lines in (C) and (D)) shows both surfaces were functionalized with $FG^{Nsp1}$-$His_{10}$ at comparable densities.

Moreover, the consistently sharp transition of the fluorescence intensity levels at the periphery of the bleached spot(s) demonstrates that the FG domains are essentially immobile and do not migrate appreciably across the surface within experimentally relevant times. It is notable that the area of contact consistently appeared darker than the surrounding when Atto488 labelled $FG^{Nsp1}$ was used (inset Figure 2B); the fluorescence though largely recovered upon separation of sphere and plane (Figure 2C-D). This suggests that the strong compression in and close to the contact area affected the fluorophore but the FG domain film remained largely stable during contact. The exact mechanism for the reduction in fluorescence is not clear. For two FG domain films (each at 5 pmol/cm$^2$) with 1 mol-% Atto488, the projected root-mean-square distance of fluorophores is 40 nm. This is much larger than the Förster distances of Atto488 (5 nm) and self-quenching is thus unlikely.

*Gap profile and contact force.* The interference of light reflected at the plane-solution and solution-sphere interfaces gives rise to a pattern of Newtonian rings. We exploited the capacity of the laser scanning microscope to acquire images of the reflected light, and such reflection interference contrast (RIC) micrographs were then analysed to quantify gap sizes, and indirectly, the contact force.

A representative RIC micrograph is shown in Figure 3A for a plane-sphere interface without a polymer interlayer. The Newtonian rings appear symmetric and without appreciable imperfections confirming that both surfaces have the expected smooth finish. The radial intensity profile (Figure 3B) could be fitted with an optical model assuming perfect plane-sphere geometry. However, the effective gap sizes at the centre of the contact thus computed were consistently negative and increased in magnitude with the applied force (Figure 3C). This indicated that there were significant deviations from the assumed ideal plane-sphere contact geometry. We hypothesized that these are due to the compressive force entailing the deformation of the planar and spherical surfaces (Figure 3D). To verify this assumption, we computed the shape of the contacting surfaces as a function of compressive force using the Hertz contact model (see Supporting Methods - *RICM analysis of a sphere pressing on a planar surface*). Subjecting the corresponding idealized theoretical RICM intensity profiles to the above-mentioned optical model indeed generated fits of good quality with negative and force-dependent effective gap sizes (Figure 3E), analogous to the experimental data. Moreover, we compared the applied forces predicted for what we operationally defined as 'soft', 'medium' and 'hard' contact in our experiments with rough estimates of the applied forces based on the magnitude of the micromanipulator's *z* motion and the mechanics of the lever arm to which the glass rod was attached (see Supporting Methods – *Estimate of compressive forces between sphere and plane*). These were in good agreement, thus demonstrating that the RIC micrographs in conjunction with the Hertz model can be exploited to estimate the contact force and to quantify the real radial gap profile as a function of the distance from the centre of contact (*i.e.* taking into account the deformation of the planar and spherical surfaces upon contact; Figure 3F).

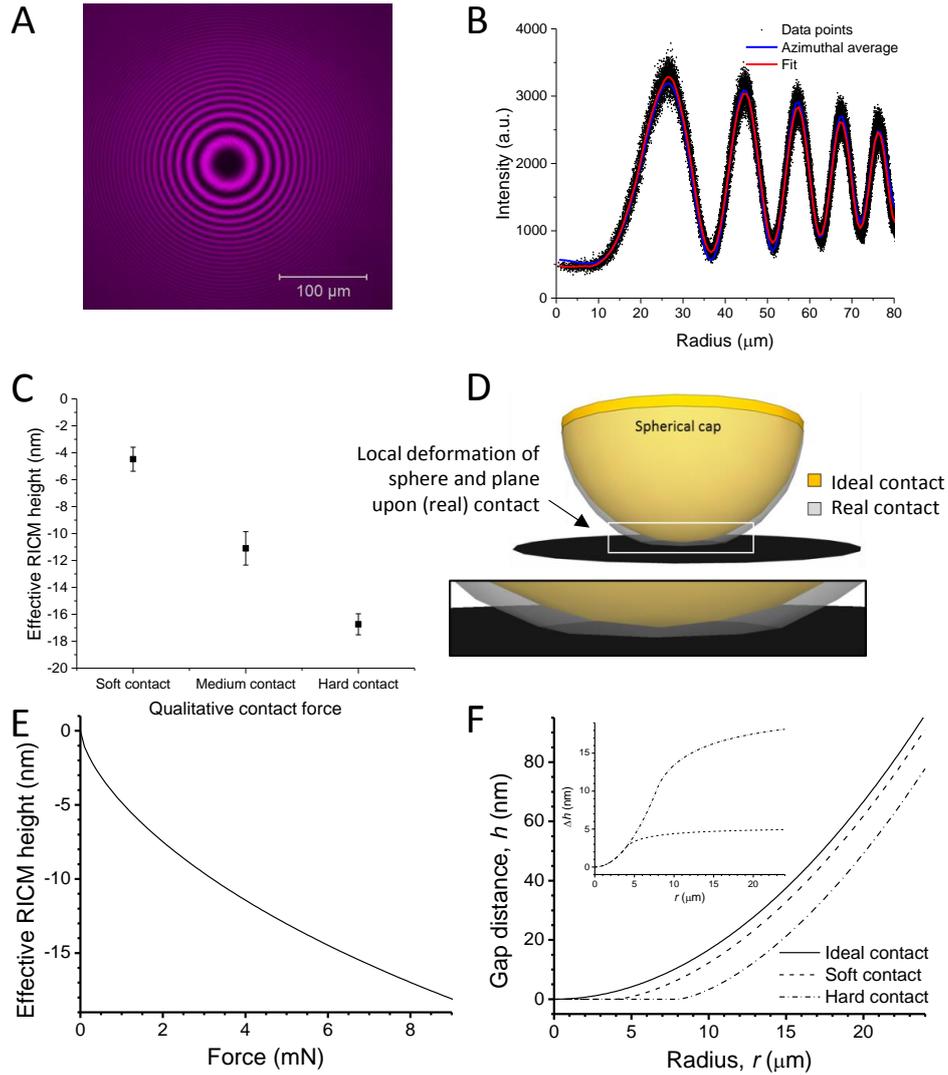

**Figure 3. Analysis of gap profiles and contact forces by RICM.** (**A**) Representative RIC micrograph of a spherical surface (glass rod) pressing on a planar surface (cover slip). Conditions: 'soft' contact, wavelength of light $\lambda = 633$ nm. (**B**) Radial intensity profile extracted from (A) (black dots), azimuthally averaged (blue line) and fitted with an optical model assuming perfect (ideal contact) plane-sphere geometry (red line). (**C**) Effective RICM height at the centre as a function of the quality of the contact, here operationally defined as 'soft' (corresponding to a few μm of micromanipulator $z$ motion following initial contact), 'medium' (~8 μm) and 'hard' (~16 μm) contact. Data points represent mean ± standard deviation of five measurements with bare surfaces. (**D**) Schematic representation of the contact geometry corresponding to an ideal plane-sphere interface (zero contact force; yellow spherical cap) and a real contact (where both surfaces are deformed at the interface owing to the finite contact force; transparent grey spherical cap). (**E**) Effective RICM height *versus* compressive force predicted from the Hertz contact model considering the geometries and mechanical properties of the glass rod and the cover slip. The curve is for two bare surfaces; if a polymer film is present between the surfaces then the effective RICM height can be increased by the optical thickness of the fully compressed polymer film to a good approximation (for details see Supporting Methods - *Estimate of compressive forces between sphere and plane*). (**F**) Radial gap profile for 'soft' ($F = 1$ mN; dashed line) and 'hard' (8 mN; dash-dotted line) contact; the idealized case of a perfect plane-sphere contact (0 mN; solid line) is also shown for comparison. The inset shows the difference in gap sizes ($\Delta h$) between the soft and hard contacts compared to the ideal contact.

The above-described validation experiments were performed with bare surfaces. RICM, however, can also be used to quantify the contact force and, subsequently, the gap profile in the presence of a polymer interlayer provided that the optical thickness of the compressed interlayer is known (see Supporting Methods - *RICM analysis of a sphere pressing on a planar surface*). From the RICM analysis, we estimate the compressive forces $F$ in our setup ranged from 1 mN at 'soft' contact to 8 mN at 'hard' contact (Figure 3C and E). It can be estimated (by considering the osmotic pressure in the FG domain film; see Supporting Methods - *FG domain film thickness under strong compression*) that forces of this magnitude would compress the FG domain film to an extent that virtually all solvent is squeezed out, essentially, leaving an incompressible polypeptide melt in the area of contact. From the $FG^{Nsp1}$ grafting density of 5 pmol/cm$^2$, the thickness of the compressed FG domain film would be 2.3 ± 0.5 nm. Considering also the presence of the EDTA surface functionalisation (which is used to graft $FG^{Nsp1}$ *via* its polyhistidine tag[33]; 0.7 ± 0.2 nm), we can estimate that this compact organic film has a thickness of $d_{compact}$ = 3.0 ± 0.7 nm (*ibid.*).

All measurements presented in the subsequent sections of the manuscript were performed at soft contact (and without any fluorescently labelled $FG^{Nsp1}$). We hence used the appropriate gap profile shown in Figure 3F whenever data for two bare surfaces were analysed, and augmented these values by 2 × 3 nm = 6 nm when the surfaces were coated with FG domain films. From the reproducibility of the compressive forces (considering the reproducibility of contact formation, Figure 3C, and also the effect of thermal drifts during data acquisition), we estimate that the gap sizes thus determined are accurate to within ± 2 nm.

**Quantification of the partitioning of macromolecules**

Having defined the polymer coating and the geometry of the confined space, we can now introduce the diffusing solute. Here, we have selected three probe molecules that have the same size but are expected to differ drastically in their interaction with FG domain films. $GFP^{Std}$ is the enhanced green fluorescent protein and is known to be weakly attracted to $FG^{Nsp1}$ through a low level of nonspecific interactions. $GFP^{Inert}$ is a mutant engineered to minimize such interactions. In contrast, $GFP^{NTR}$ is a mutant engineered to gain properties much alike a nuclear transport receptor with an enhanced attraction to $FG^{Nsp1}$. These probe molecules originate from a recent study where the surface features of GFP were explored with respect to its NPC-translocation rate.[37] Overall, a distinct correlation between NPC-passage rate and partitioning into macroscopic FG domain hydrogels was observed in these assays. In ref. [37], $GFP^{Inert}$ is called SinGFP4A, and $GFP^{NTR}$ is called 7B3.

In a first instance, we focused on the distribution of GFP variants in the $FG^{Nsp1}$ films. Figure 4A and B show fluorescence micrographs of $GFP^{Std}$ surrounding the contact point between the planar and the hemi-spherical surface (centre of image), for bare and $FG^{Nsp1}$ functionalized surfaces, respectively. Although this is not immediately apparent in the micrographs, the corresponding radial intensity profiles clearly reveal that GFP$^{Std}$ was partly excluded from the FG domain film (Figure 4E). Equivalent micrographs of GFP$^{Inert}$ and GFP$^{NTR}$ surrounding the contact point between FG domain functionalized surfaces are shown in Figure 4C and D, respectively. Similar to GFP$^{Std}$, GFP$^{Inert}$ was also excluded from the FG domain film albeit to a greater extent. In contrast, as evident from the micrographs and the corresponding radial intensity profiles, GFP$^{NTR}$ was substantially enriched in the FG domain film. Supporting Figure S3 shows further controls for the specificity of the FG$^{Nsp1}$ film interactions with the used GFP variants.

For further analysis, we focused on the confined region in which the FG domain films that coat the planar and spherical surfaces overlap. Based on the geometry of the confined space (established by RICM for 'soft' contact as shown in Figure 3F, and the additional $2 \times 2.3$ nm = 4.6 nm of the compacted FG$^{Nsp1}$ film in the contact area) and a thickness of ~30 nm per uncompressed FG$^{Nsp1}$ film (Supporting Figure S1) plus $2 \times 0.7$ nm = 1.4 nm for the APTES functionalisation, one can estimate that this zone extends 19 µm from the centre of the contact area.

Because the extension of the confocal volume in $z$ is much larger than the gap size, the intensities shown in Figure 4E can be expected to scale with the areal density of GFP molecules (*i.e.* GFP molecules per unit of projected area). Thus, by re-scaling the intensity by the gap size, a measure of the GFP concentration within the gap volume can obtained. This data is shown in Figure 4F as a function of the gap size. In the case of GFP$^{Std}$ confined between bare surfaces the re-scaled intensity is constant for gap sizes of 20 nm and more; it gradually decreases towards smaller distances and practically attains zero around 5 nm. The observed trends are consistent with expectations for simple volume exclusion: GFP has a size of 5 nm and should thus not penetrate into gaps smaller than that, and depletion effects at the wall are expected to lead to a gradual increase in concentration for small gap sizes until a plateau corresponding to the bulk concentration is effectively reached. The match with these expectations lends support to the validity of the analytical approach. In addition, this control has the benefit of enabling conversion of the re-scaled intensities into concentrations: by identifying the plateau value of $I_{\text{re-scaled}} = 3.5$ with the bulk concentration $c_{\text{bulk}} = 2$ µM, we have $c = 2$ µM / $3.5 \times I_{\text{re-scaled}}$.

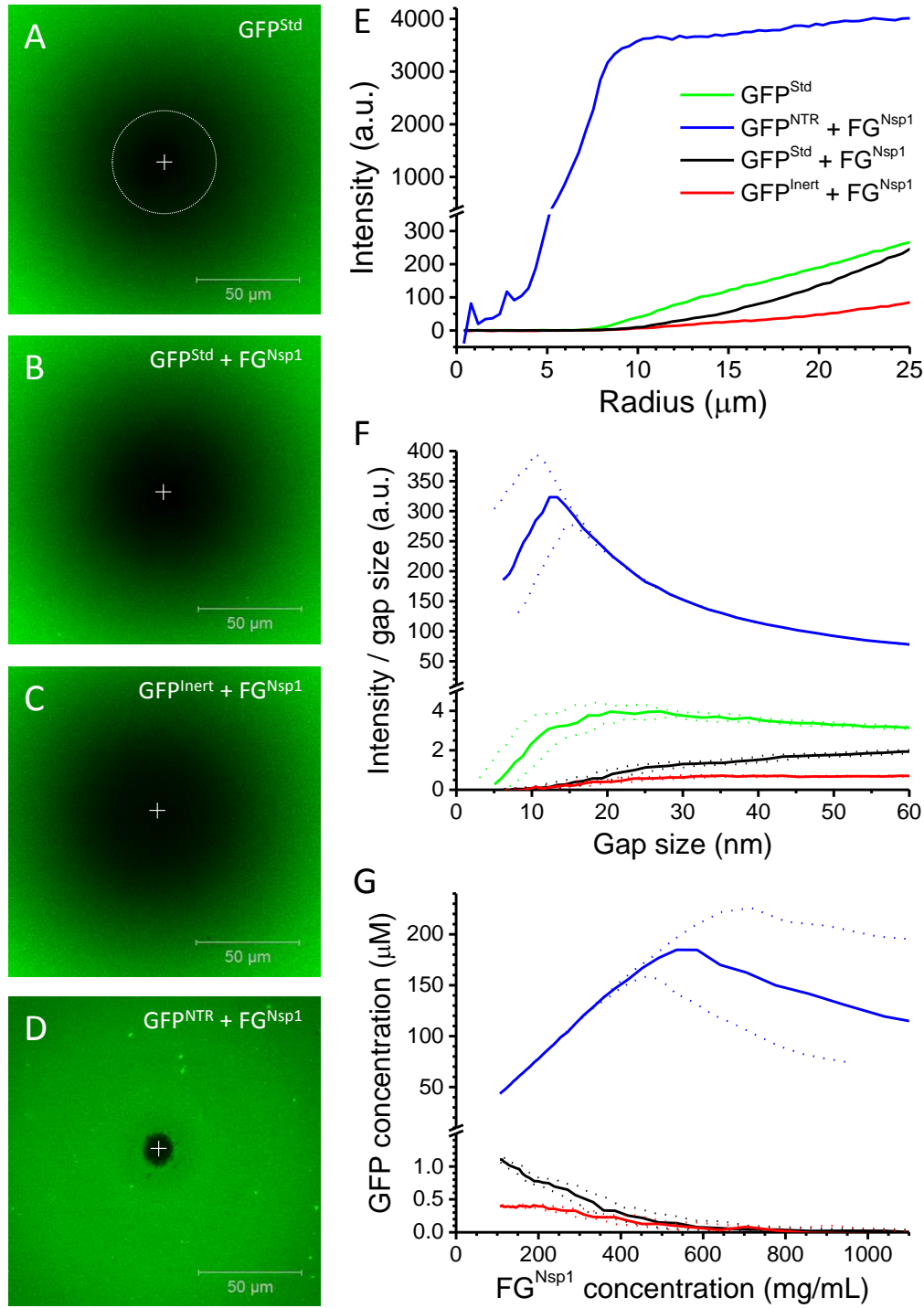

**Figure 4. Analysis of GFP distribution inside the polymer film with fluorescence microscopy. (A)** Fluorescence micrograph of GFP$^{Std}$ surrounding the contact area (the white cross indicates the location of centre and the diameter of contact area) between bare planar and spherical surfaces. **(B)** As in (A) but with both surfaces functionalized with a FG$^{Nsp1}$ film. **(C-D)** As in (B) but with GFP$^{Inert}$ (C) and GFP$^{NTR}$ (D) instead of GFP$^{Std}$. **(E)** Integrated radial intensity profiles derived from the micrographs in (A-D). The white dotted circle in (A) illustrates the area analysed (the radius was measured from the centre of the contact area). **(F)** Intensity profiles of micrographs in (A-D), normalized with the gap size between the planar and the spherical surface, plotted *versus* the gap size (upper curve for each sample). Solid lines were computed with the most probable gap size; dotted lines delineate the confidence interval based on the estimated ± 2 nm uncertainty in gap size. A gap size of 60 nm here occurs at a radius of approximately 20 μm. **(G)** Same as (F) but recalculated to GFP concentration and plotted *versus* the FG$^{Nsp1}$ concentration.

In FG$^{Nsp1}$ films, all GFP variants show a behaviour that differs from GFP$^{Std}$ between bare surfaces: GFP$^{Std}$ and even more so GFP$^{inert}$ are depleted, whereas GFP$^{NTR}$ is strongly enriched. Moreover, it is notable that the concentration of all GFP variants varies substantially with gap size. Figure 4G shows the same data as Figure 4F but with the gap size converted to FG$^{Nsp1}$ concentrations based on the known areal mass density of 320 ng/cm$^2$ (corresponding to 5 pmol/cm$^2$, or a root-mean-square distance between anchor points of approximately 6 nm; Supporting Figure S1) for each of the two apposed FG$^{Nsp1}$ films. This plot represents the first main outcome of the PSCM method. A notable finding is that the concentration of GFP$^{NTR}$ increases with FG$^{Nsp1}$ concentration (with a linear dependence) over a substantial range of FG domain concentrations (from ~100 for the uncompressed film to ~500 mg/mL) before it shows the decrease that can be consistently seen for GFP$^{inert}$ and GFP$^{Std}$. We note here that the end of the concentration scale in Figure 4G (1.2 mg/mL) is already very close to a solvent-free polypeptide 'melt' (density 1.4 mg/mL).

The partition coefficients (Figure 5), describing the partitioning of probe molecules between bulk solution and the FG domain1 film, were determined from the data presented in Figure 4F by calculating the ratios of intensities for GFP$^{Std/NTR/Inert}$ in FG$^{Nsp1}$ films and GFP$^{Std}$ between bare surfaces. Figure 5A shows how the partition coefficient varies within the overlapping FG domain films. It was evident that GFP$^{Std}$ and GFP$^{Inert}$ were excluded from the FG$^{Nsp1}$ film (partition coefficients < 1) while GFP$^{NTR}$ was strongly enriched. Figure 5B illustrates how this differential effect is substantially enhanced when the FG domain film is compressed and thus more concentrated.

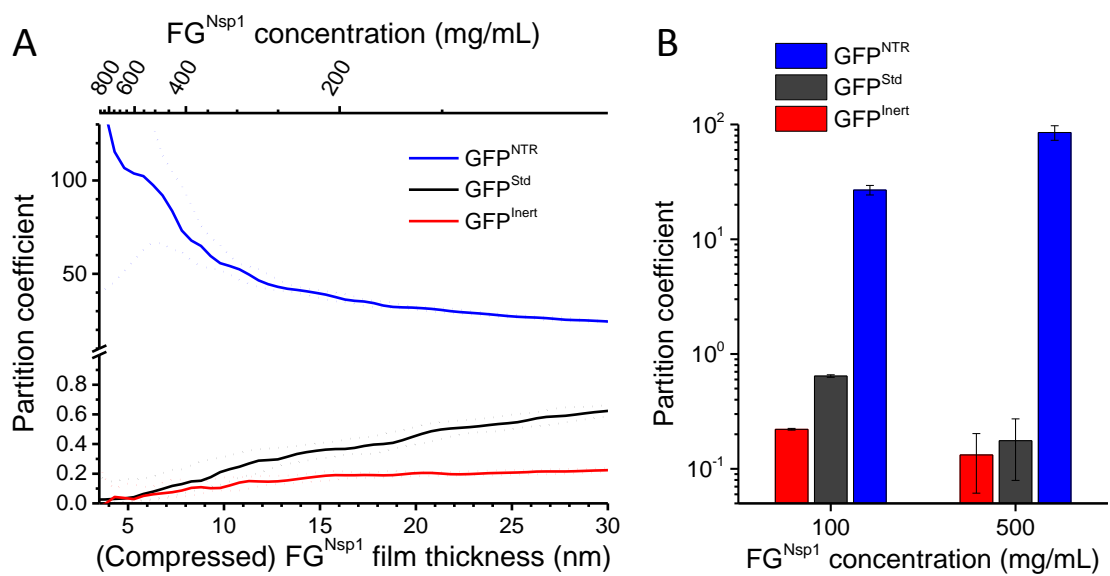

**Figure 5. Analysis of GFP partitioning inside the polymer film with fluorescence microscopy. (A)** Partition coefficients of GFP$^{Std}$, GFP$^{NTR}$ and GFP$^{Inert}$ inside the FG$^{Nsp1}$ film, calculated from the data in Fig. 4F as a function of the FG$^{Nsp1}$ concentration and (compressed) film thickness. **(B)** Comparison of partition coefficients for FG$^{Nsp1}$ concentrations of approximately 100 and 500 mg/mL, corresponding to a virtually uncompressed (30 nm thick) and strongly compressed (6 nm thick) FG$^{Nsp1}$ film, respectively. Mean values from two independent measurements per GFP variant are shown; error bars represent highest and lowest values obtained.

**Quantification of macromolecular diffusion within confined polymer layers**

In contrast to conventional FRAP, line FRAP enables the analysis of spatial variations (*i.e.* along the bleached line) in diffusion in a single measurement with a resolution down to a few micrometres. We chose this approach as it is particularly well suited to probe how the diffusion varies with the gap size, and thus, the polymer film thickness and concentration.

The kymograph in Figure 6A shows a line FRAP dataset for GFP$^{NTR}$ in FG$^{Nsp1}$ films, where the imaged line was set to go through the centre of the plane-sphere interface. The photo-bleached part of the line (cutting asymmetrically across the centre) and the subsequent fluorescence recovery are readily visible in this crude presentation and demonstrate that GFP$^{NTR}$ is mobile everywhere in the confined area except in the ~10 µm wide central exclusion zone which is hardly penetrated. Figure 6B shows a recovery curve obtained by averaging over a 3 µm wide section of the line (encased in white in Figure 6A). The best fit with a diffusion model (red line) assuming a mobile fraction $k$ with diffusion coefficient $D$ reproduces the data well and confirms that the vast majority of GFP$^{NTR}$ is mobile ($k = 0.87 \pm 0.01$; note that equilibrium is not reached within the measured recovery phase of 1.2 s). A possible explanation for the small fraction of apparently immobile GFP$^{NTR}$ ($1 - k = 0.13 \pm 0.01$) may be residual non-specific interactions of the protein with the surfaces (Supporting Figure S3).

Performing such analyses along the bleached line reveals how the diffusion constant varies with the distance from the centre and thus, with the gap size or polymer concentration. Figure 6C illustrates how the diffusion coefficient of GFP$^{NTR}$ varies with the distance from the centre based on Figure 6A. Figure 6D shows how the GFP$^{NTR}$ diffusion coefficient (averaged from multiple measurements), varies with FG domain film thickness and concentration. Reassuringly, the mobile fraction was consistently high across the full FG domain thickness range (and all measurements) at $k = 0.84 \pm 0.02$ (inset in Figure 6D), suggesting that possible surface effects do not skew the diffusion data appreciably.

In Figure 6D-E, it can be seen that GFP$^{NTR}$ diffuses with $D = 1.6 \pm 0.2$ µm$^2$/s at the point where the FG domain films just overlap (film thickness ≈ 30 nm; FG$^{Nsp1}$ concentration ≈ 100 mg/mL), and that the diffusion constant decreased only moderately with increasing film compression and concentration. From analogous measurements with GFP$^{Std}$ (Supporting Figure S4) we estimate $D = 6.5 \pm 1.9$ µm$^2$/s for the unperturbed FG domain film. For comparison, the diffusion coefficient of GFP in aqueous solution has been determined by fluorescence correlation spectroscopy to $D = 90 \pm 3$ µm$^2$/s.[38] Thus, the FG$^{Nsp1}$ film reduces the diffusion of GFP$^{Std}$ (and likely also GFP$^{Inert}$) by about an order of magnitude, whilst GFP$^{NTR}$ experiences a further reduction by a moderate few fold.

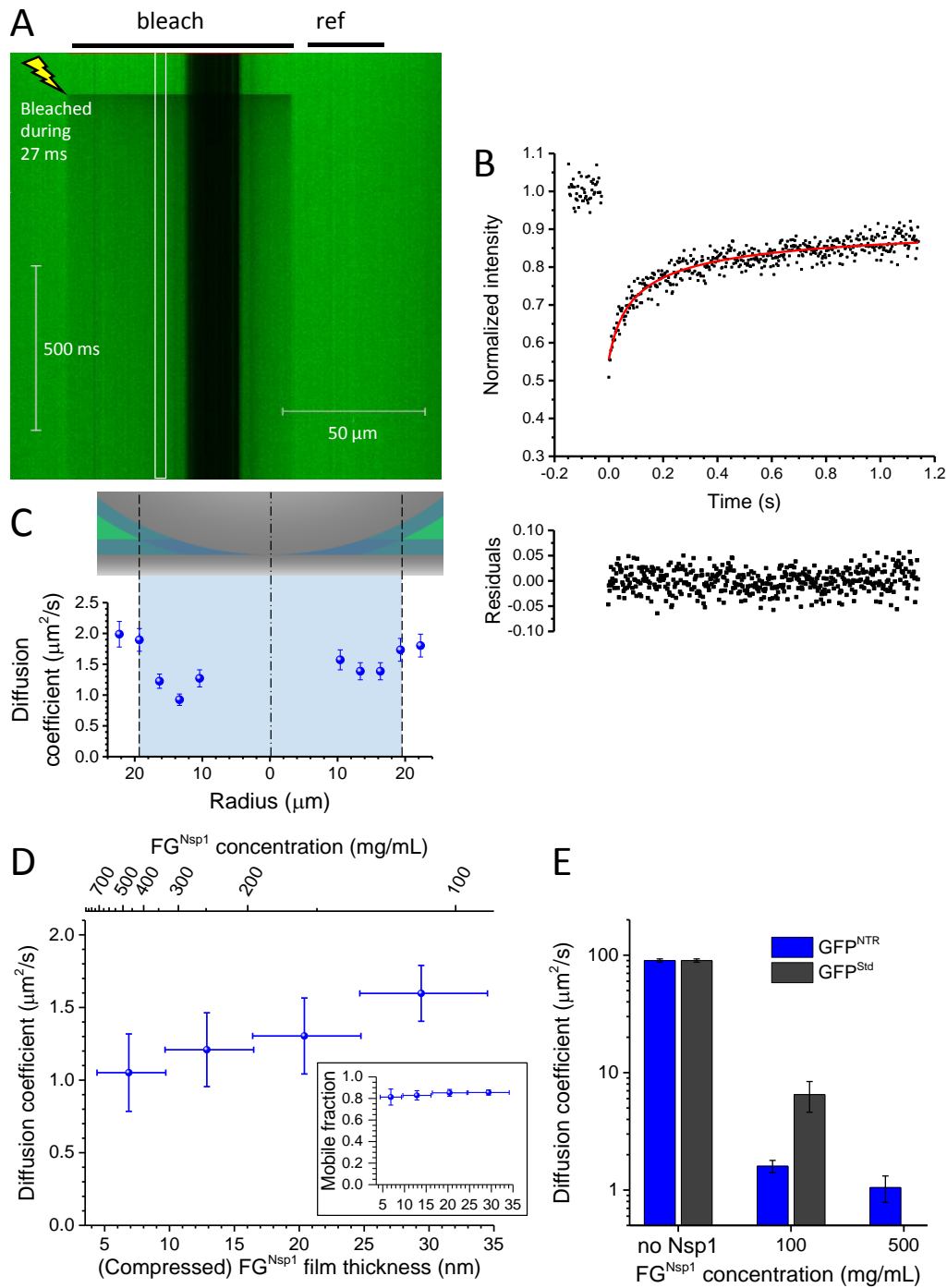

**Figure 6. Analysis of macromolecular diffusion along confined polymer films by line FRAP.** (A-D) Representative data for GFP[NTR] in an FG[Nsp1] film to illustrate the data acquisition and analysis. **(A)** Kymograph of a scan line across the centre of the plane-sphere contact area (*cf.* Figure 4D). The two lines on top mark parts of the scan line that are photo-bleached ('Bleach'; yellow flash marks time point of bleaching) and used as reference to correct for bleaching during imaging ('Ref'), respectively. **(B)** Fluorescence recovery curve (black dots) obtained from the data encased with a white box in (A). The best fit with the line FRAP model (red line; residuals from the fit are shown below) gives $D = 1.4 \pm 0.2$ $\mu m^2/s$, $k = 0.87 \pm 0.01$ and $K_0 = 0.88 \pm 0.03$. **(C)** GFP[NTR] diffusion coefficients at various distances from the contact point; error bars represent the standard error of the fit. **(D)** GFP[NTR] diffusion coefficient (main panel) and mobile fraction (inset) as a function of the FG[Nsp1] concentration and (compressed) film thickness. Mean and standard deviations for 8 data points are shown (2 data points per image, left and right of the centre, from a total of 4 images selected from 2 independent measurements); mobile fraction

and bleaching parameter across these measurements were roughly constant: $k = 0.84 \pm 0.02$ and $K_0 = 0.93 \pm 0.05$. **(E)** Comparison, for $GFP^{NTR}$ and $GFP^{Std}$, of diffusion constants in bulk solution ('no $FG^{Nsp1}$'; taken from ref. [38]) and for $FG^{Nsp1}$ concentrations of approximately 100 and 500 mg/mL, corresponding to a virtually uncompressed (30 nm thick) and strongly compressed (6 nm thick) $FG^{Nsp1}$ film, respectively.

## DISCUSSION

**Salient performance features of PSCM**

We have demonstrated that PSCM provides a radial gap profile (by RICM, with an accuracy in the gap size of a few nm; Figure 3), a radial solute concentration profile (by fluorescence microscopy; Figure 4E) and a radial solute diffusion profile (by line FRAP; Figure 6C). These data can be correlated for each radial position, and thus PSCM enables quantitation of a wealth of information about the interaction of solutes with solvated polymer films in a single experiment. For polymer films of known thickness and/or surface coverage, the gap profile can be readily translated into a film compression profile and a polymer concentration profile, respectively. Partition coefficients can thus be obtained not only between the bulk and the uncompressed polymer phase, but also, as a function of the compression and concentration of the polymer phase (Figure 5). Importantly, in-plane diffusion becomes quantifiable with PSCM for the uncompressed polymer phase and as a function of the compression and concentration of the polymer phase (Figure 6D-E).

Several extensions to the presented capabilities of the PSCM method are conceivable. The determination of contact forces, and ultimately gap profiles, required the optical thickness of the fully compressed, solvent free polymer interlayer to be determined with other methods. For our FG domain films, we used a combination of QCM-D and spectroscopic ellipsometry (see Supporting Figure S1 and Supporting Methods), though other techniques are also available. Alternatively one can quantify the optical thickness of the polymer interlayer using the RICM capability of PSCM, if the contact force is controlled by other means. A defined and constant contact force may be realised with some form of force balance, such as gravitation, for example. This enables the gap profile to be accurately determined for further analysis of solute diffusion and partitioning in less well characterised polymer interlayers (for details, see Supporting Methods - *RICM analysis of a sphere pressing on a planar surface*). Moreover, contact forces can be adjusted depending on the requirements of the polymer film of interest: for resilient films they can be made large enough such that essentially all solvent is being squeezed out in the contact area for a maximal range of compression to be probed; for fragile films forces can be kept small enough to avoid excessive damage.

In Figure 4G and Figure 5A we demonstrated that solute binding can be quantified as a function of polymer concentration in a single experiment. If such data are additionally acquired for a set of solute concentrations, then it becomes possible to obtain binding 'isotherms' as a function of polymer concentration in a single experiment. Whilst such data can also be obtained by other means,[25] PSCM can provide them with higher throughput. Moreover, PSCM can uniquely probe in-film diffusion as a function of solute and polymer concentration. This not only enables the concentration dependent diffusivity of the solute to be quantified, but may also be exploited to measure how solutes affect the diffusivity of (fluorescently labelled) components of the polymer film itself.

For some applications, the possibility of determining the partition coefficient and/or diffusion constant of a molecule or nanoparticle within an uncompressed film may be particularly attractive. This requires a precise knowledge of film thickness to determine where exactly along the sphere-plane gap profile free solvent is excluded while polymers remain uncompressed. Our results (Figure 5A and Figure 6D) show that the measured values changes gradually when transitioning from a compressed film (film thickness < 30 nm) to an uncompressed film with some free solvent (> 30 nm). Hence, an approximate knowledge of the film thickness is sufficient to obtain good estimates of the partition coefficient and diffusion constant within an uncompressed FG domain films. Other polymer films and solutes may though present very different partitioning and diffusion profiles; with a sharper transition it may become possible to infer the film thickness from the partitioning or diffusion profiles.

The presented confinement technique is versatile. It is compatible with most confocal and epi-fluorescence microscopy setups, and can readily be added onto old or new microscopes. It is also compatible with all common methods to measure diffusion such as FRAP (including line FRAP), FCS (including raster image correlation spectroscopy, RICS[39]) and SPT. For fluorescence-based SPT, conventional experiments usually require total internal reflection illumination (TIRF) to reduce background fluorescence signal from the bulk solution.[40] With PSCM, a good signal to noise ratio can be expected even without TIRF because the confinement already effectively avoids background. This simplifies experiments, and for gap sizes smaller than ~100 nm it is even more effective than TIRF. Also, with three-dimensional SPT, it would be possible to analyse in-plane diffusion as well as out-of-plane diffusion and hence to probe diffusion anisotropy in polymer films. Last but not least, the confinement technique should also be compatible with specialized non-fluorescent imaging and particle tracking modalities (*e.g.* photothermal microscopy[41]).

**New insights into NTR-FG domain interactions, and functional implications for NPC permselectivity**

In addition to establishing PSCM, this study also provided new insights into the dynamics of NTR-FG domain interactions.

*The FG domain concentration differentially affects uptake of NTRs and inert macromolecules.* NTR binding depends on FG domain concentration in a non-monotonous way. For the model NTR used here (GFP$^{NTR}$), maximal binding occurred around 500 mg/ml (Figure 4G), a concentration that likely exceeds the FG domain concentration in the nuclear pore. In previous work, we had already found circumstantial evidence for such a non-monotonous dependence for the NTR NTF2 in films made of an artificial, regular repeat of FSFG motifs.[25] Collectively these data suggest that a non-monotonous dependence is a common phenomenon, although further experiments will be required to quantify how this depends on NTR and FG domain types. We had previously shown that NTR binding to FG domain assemblies is determined by a balance of attractive interactions of NTRs with FG motifs and excluded volume repulsion.[25] Whilst both types of interaction can be expected to increase with FG domain density, our data suggest that the increase in attractive interactions dominates at low and intermediate FG domain concentrations (up to several 100 mg/mL) whereas excluded volume repulsion take over at the highest FG domain concentrations, thus giving rise to a non-trivial concentration dependence. For inert macromolecules, on the other hand, attractive interactions are minimal and uptake should decrease monotonously with FG domain concentration. This is indeed clearly evident for $GFP^{Std}$ and $GFP^{Inert}$ (Figure 5A).

The opposite effects of FG domain concentration on the uptake of NTRs and inert macromolecules is intriguing, as it implies that there exists an optimal FG domain concentration where the NTR uptake is the most selective. In our specific experimental case we can define selectivity of uptake as the ratio of partition coefficients, and see that the selectivity of $GFP^{NTR}$ over $GFP^{inert}$ is 27 / 0.22 ≈ 120 at 100 mg/mL $FG^{Nsp1}$ (*i.e.* for the uncompressed FG domain film of 30 nm thickness) and 85 / 0.13 ≈ 650 at 500 mg/mL $FG^{Nsp1}$ (when the FG domain film is compressed to 6 nm; Figure 5B). We note here that even more dramatic selectivity values have been reported for microphases of other FG domains,[37] and for real NTRs with $FG^{Nsp1}$ films;[3,25] this however may arise at least to some extent because either the FG domain (in the microphases) or NTR (in $FG^{Nsp1}$ films) were different. FG domains are known to exhibit a certain level of cohesiveness, which promotes the formation and determines the properties of FG domain phases[34,42-45], and is also essential for the formation of a functional permeability barrier[27,44]. Phases of the most cohesive natural FG domains indeed exhibit a rather high FG domain concentration (several 100 mg/mL) yet still retain a significant amount of solvent.[34,43] We propose an enhanced selectivity of NTR uptake as a novel, previously unrecognised, benefit of FG domain cohesiveness. It should be noted that the level of cohesiveness has to be balanced not only to maximise selectivity of NTR uptake but also because excessive cohesiveness may induce phase separation at the nanoscale and thus an effective breakdown of the permselectivity barrier in the NPC, as reported previously.[34]

*FG domain phases slow down NTR diffusion only moderately.* $GFP^{NTR}$ diffusion in $FG^{Nsp1}$ films depends only weakly on the $FG^{Nsp1}$ concentration, and the diffusion rate is not much lower than that of GFP in the cytosol ($D = 6.1 \pm 2.4$ µm$^2$/s for the cytoplasm in *E. coli*[46]). This finding is consistent with a moderate reduction in diffusion inside the NPC (as compared to the cytoplasm) for an import complex made from the NTRs importin α and β and a GFP dimer model cargo.[29] Future tests with other NTRs and FG domains can show if this is generally true. If so, this would reflect a remarkable adaption of NTR-FG domain interactions to the function of NTRs: enrichment in the nuclear pore which is beneficial to transport but requires strong interactions with FG domains is accomplished without a significant penalty on diffusion (which is generally slowed down by the attractive interactions). Most likely this is a consequence of the interactions of NTRs with individual FG motifs being very fast.[31] With PSCM and designer FG domains and NTRs it now becomes possible to probe experimentally how NTR diffusion is defined by the multivalent nature of the interaction between NTRs and FG domains.

To estimate the magnitude of the effects that diffusion and partitioning in the FG domain phase have on fluxes $J$ across the NPC, we consider the simple theoretical model of the steady-state flux by Frey and Görlich,[42] who arrived at $J = ADk_{entry}\Delta c/(Lk_{exit} + 2D)$, where $A$ and $L$ are the effective cross section and length of the NPC channel, respectively, $D$ is the diffusion constant inside the FG domain phase, $k_{entry}$ and $k_{exit}$ are the rate constants for entering and exiting the channel, and $\Delta c$ is the concentration difference across the channel. With the partition coefficient $P = k_{entry}/k_{exit}$, this equation can be recast into $J = A\Delta c/(LD^{-1}P^{-1} + 2k_{entry}^{-1})$. If fluxes are limited by the diffusion through and exit from the pore ($k_{entry} \gg 2DP/L$), then $J \propto DP$. Under this condition, any moderate decrease in $D$ is overcompensated by a much larger increase in $P$, leading to an enhanced flux of NTRs compared to similar-sized inert molecules. Taking our results for GFP$^{NTR}$ and GFP$^{inert}$ at 100 mg/mL FG$^{Nsp1}$ as an example, we have a diffusion constant ratio of 1.6 µm$^2$/s / 6.5 µm$^2$/s ≈ 0.25, a partition coefficient ratio of 27 / 0.22 ≈ 120, and thus a 30-fold enhanced flux of GFP$^{NTR}$ over GFP$^{inert}$. If instead entry into the pore is rate limiting ($k_{entry} \ll 2DP/L$), then $J$ does not depend on $D$ or $P$, and differences in flux instead arise from a larger entry rate of NTRs over inert molecules (not quantitated here).

The diffusion of GFP$^{Std}$ in the FG$^{Nsp1}$ film is moderately reduced (by about an order of magnitude) compared to the bulk solution. This implies that the correlation length ('mesh size') within the FG$^{Nsp1}$ film must be close to the size of GFP (cylinder with 4.2 nm length and 2.4 nm diameter).[34] This is indeed quite reasonable considering the grafting density and volume density of the FG$^{Nsp1}$ film. It is also consistent with a moderate level of GFP$^{Std}$ and GFP$^{Inert}$ exclusion from the FG$^{Nsp1}$ film (Figure 5B). For inert macromolecules that are significantly larger than the mesh size, polymer theory predicts the diffusion (and uptake) to be much reduced.[47,48] PSCM now provides a tool to quantitate these effects and test the theoretical predictions for FG domain assemblies of defined composition and concentration.

**Concluding remarks**

In summary, we have presented an analytical method that allows quantitative characterization of macromolecular diffusion within (10s to 100s of nanometre) thin solvated polymer coatings. The method can be integrated with conventional optical microscopes, and is versatile. It provides quantitative information about the diffusion of macromolecules within the polymer coating, and about the partitioning of macromolecules between the polymer film and the bulk solution. Thanks to the shape of the confined geometry, these parameters can also be mapped as a function of polymer film compression (and concentration) in a single experiment. The described methodology is generic and may find widespread use in the analysis of solvated polymer films and their interaction with fluorescent macromolecular probes. An obvious application in basic science are biomimetic model systems (*e.g.* for the nuclear pore permselectivity barrier, as presented here), where this method can provide new insight into transport processes in complex polymeric environments. However, the potential use is much broader and the methodology should find use in the development of functional coatings for a wide range of applications, from fundamental research in polymer and biological physics to everyday-life applications in biomaterials and paints.

Using the case of the nuclear pore permselectivity barrier we demonstrate direct quantitation of the diffusion coefficient of an NTR-like molecule within nanoscale assemblies of FG nucleoporins, and demonstrate that the FG domain concentration sensitively affects the selectivity of NTR uptake. This data opens up avenues for further investigations to understand the physical mechanism underpinning the exquisite permselectivity of the nuclear pore complex.

## MATERIAL AND METHODS

**Materials**

Chemicals were obtained from commercial sources and used without further purification. Ultrapure water (resistivity 18.2 MΩ/cm) was used throughout. The FG domain of Nsp1 (amino acids 2 to 601) from *S. cerevisiae* with a C-terminal $His_{10}$ tag ($FG^{Nsp1}$-$His_{10}$; 64.1 kDa) was produced and purified as described earlier.[3,34] For fluorescent labelling, the N-terminal cysteine of $FG^{Nsp1}$-$His_{10}$ was reacted with Atto488-maleimide as described previously.[42] $FG^{Nsp1}$-$His_{10}$ variants were stored at concentrations between 11.5 and 15.6 μM (7.4 and 10 mg/ml) in 50 mM Tris, pH 8, supplemented with 6 M guanidine hydrochloride (GuHCl) at -80 °C. Before use, the FG domains were diluted in working buffer (10 mM Hepes, 150 mM NaCl, pH 7.4) to a final concentration of 0.16 μM (0.1 mg/ml).

Three probe molecules derived from green fluorescent protein (GFP) were used. $GFP^{Std}$ is the well known enhanced GFP. $GFP^{NTR}$ and $GFP^{Inert}$ are mutants that are described in detail in ref. [37]. $GFP^{NTR}$ (denoted 7B3 in ref. [37]) exhibits the qualities of an NTR in terms of facilitated transport through nuclear pores and in macroscopic FG domain hydrogels. $GFP^{Inert}$ (denoted SinGFP4A in ref. [37]) is 'superinert' and is effectively excluded from nuclear pores and macroscopic FG domain hydrogels. Before use, the GFP samples were diluted in working buffer, to a final concentration of 2 μM unless otherwise stated.

Glass cover slips (24 × 24 $mm^2$, #1.5, made from Schott D 263 M glass) were purchased from Thermo Scientific. Rods of borosilicate glass (type 1 class A) with a diameter of 5 mm were purchased from VWR. These were cut into 25 mm long pieces and ends polished to approximately hemi-spherical caps with a radius of curvature of approximately 3 mm. The surfaces thus prepared were smooth on the nanometre scale, with a root mean square roughness of 0.4 nm as measured by atomic force microscopy (Supporting Figure S5A).

**EDTA functionalization of glass surfaces**

Glass cover slips and glass rods with hemi-spherical caps to be functionalized with FG domains were pre-functionalized with EDTA, according to an established procedure,[33] to allow binding of polyhistidine tagged proteins. Initially, the surfaces were cleaned by 10 min sonication in 2% SDS and water, respectively. After rinsing with water, surfaces were first blow dried using nitrogen gas ($N_2$) and then treated with UV/ozone (ProCleaner 220, BioForce Nanosciences, USA) for 30 min. A desiccator harbouring 30 µl APTES (without any solvent) was purged with $N_2$ for 2 min. The glass surfaces were then placed inside, followed by purging with $N_2$ for another 3 min. The desiccator was sealed and the surfaces were incubated for 1 h. The surfaces were then sequentially incubated in four freshly prepared aqueous coupling solutions (0.5 M EDTA, 0.25 mM EDC, pH 8.0), once for 3.25 h, twice for 2 h and then once for 15 h. After the final incubation the surfaces were rinsed with water and blow dried with $N_2$. This surface coating did not enhance the surface roughness appreciably (Supporting Figure S5B). The EDTA functionalised surfaces were stored in air at room temperature until use.

**Optical microscopy and setup of the plane-sphere confinement microscopy (PSCM)**

All microscopy experiments were performed using an inverted laser scanning microscope (LSM 880; Zeiss, Oberkochen, Germany) equipped with a 40× oil immersion objective having a numerical aperture of 1.4 (Plan-Apochromat 40x/1.4 Oil DIC M27). Images of 512 × 512 or 1024 × 1024 pixels were captured using a pixel dwell time of 2.06 µs. For fluorescence imaging the pinhole size was set to 5 airy units. This setting provided for robust alignment of the mid-plane of the confocal volume with the plane-sphere interface at a suitable lateral resolution ($r_{xy} = 0.50 \pm 0.04$ µm determined experimentally at 488 nm laser wavelength; Supporting Figure S6).

The sample chamber consisted of a custom-made PTFE holder to the planar bottom of which a suitably functionalized glass cover slip was attached using silicon glue (Twinsil; Picodent, Wipperfürth, Germany). The holder with coverslip was then mounted on the microscope stage. They formed the walls of a cylindrical cuvette of 10 mm diameter, the axis of which was coarsely aligned with the optical axis.

To form FG domain films, EDTA-functionalized planar and hemi-spherical surfaces were incubated first with 2 mM $NiCl_2$ in working buffer (15 min) and then with 0.16 µM (0.1 mg/mL) $FG^{Nsp1}$-$His_{10}$ in working buffer (30 min). After the latter incubation step, excess sample was removed by serial dilutions with working buffer. To visualize the FG domain film, 1 mol-% of fluorescently labelled $FG^{Nsp1}$-$His_{10}$ was mixed into the $FG^{Nsp1}$-$His_{10}$ solution in some experiments. Throughout the experiment, protein-coated surfaces were kept in working buffer to prevent drying. Probe molecules were added to reach a final concentration of 2 µM unless otherwise stated.

One end of a suitably functionalized rod with hemi-spherical caps was lifted into the cylindrical cuvette with the aid of a micromanipulator (PatchStar; Scientifica, Uckfield, United Kingdom). Transmitted and reflected laser light served as guidance to facilitate coarse and fine alignment, respectively, of the rod axis with the optical axis before the spherical cap and the planar cover slip were brought into contact (Supporting Figure S2). Once contact between the surfaces was reached, the area of contact and its close surroundings (typically 150 × 150 µm$^2$) were imaged. In addition, RICM and FRAP experiments were carried out as described below.

The background fluorescence intensity was recorded with the focus position set 50 µm below the solid-liquid interface of the planar surface, *i.e.* within the glass coverslip. The average fluorescence intensity of such images was determined using ImageJ software.

**Reflection interference contrast microscopy (RICM)**

The plane-sphere geometry allows for the use of RICM, a well-established technique[49] that utilizes the interference pattern created by reflections at the apposed planar and curved interfaces to determine the gap profile between them. In conventional RICM applications, the typical size of the spherical probe is in the micrometre range, and RICM has previously been combined with colloidal probe atomic force microscopy, to study the mechanical properties of polymer brushes.[50] In contrast, the hemi-spherical cap used in our setup has a radius in the millimetre range. Therefore, for RICM imaging the pinhole was opened to the maximum and the focus was positioned a few micrometres below the upper surface of the planar glass cover slip. This provided a high contrast image of the circular interference pattern (Newtonian rings) with minimal stray light. RIC micrographs with the interference patterns were analysed with a custom written algorithm implemented in LabView (described previously[51]) to quantify the effective height at the centre of the plane-sphere interface. Data were fitted over an area of 170 × 170 µm$^2$ typically encompassing six full interference fringe rings. Fixed input parameters for the algorithm were the radius of curvature of the hemi-spherical cap ($R$ = 3 mm), the wavelength ($\lambda$ = 633 nm), the pixel size, the refractive index of the buffer ($n$ = 1.334), and the illumination numerical aperture (INA = 0.999). The INA was experimentally determined by imaging the interference fringes formed between two non-parallel cover slips (air wedge in between the two surfaces) and by subsequently fitting the obtained intensity profile as described by Rädler *et al*.[52] Adjustable parameters in the fitting routine were the effective RICM height along with two parameters accounting for background intensity, two parameters for amplitude normalization and one parameter accounting for residual defocus and errors in $R$.

**Fluorescence recovery after photobleaching in line mode (line FRAP)**

The diffusion of probe molecules within the overlapping polymer brushes was quantified using line FRAP.[53] A single line, across the point of contact between the two surfaces, was imaged continuously. After a number of scans, a part of the line was bleached and the fluorescence recovery was then monitored.

Kymographs for line FRAP analysis were acquired as a times series of 512 line scans over 512 pixels. The first 50 line scans were used to acquire pre-bleach data. A selected part (260 pixels) of the line was then bleached using the maximal intensity of the 488 nm laser (10 bleach iterations; total bleach time of 27.2 ms), and the remaining lines were used to monitor the fluorescence recovery. Fluorescence recovery profiles were extracted from the kymographs using ImageJ and fits with the line FRAP equation were performed in Origin Pro (OriginLab, Northampton, MA, USA).

The normalized fluorescence intensity was determined by $I_{norm}(x,t) = (I_{meas}(x,t) - I_{bg}) / (I_{pre}(x) - I_{bg})$, where $I_{meas}(x,t)$ is the measured intensity at position $x$ and time $t$ after bleaching, $I_{bg}$ is the mean background intensity (measured by focusing inside the glass cover slip), and $I_{pre}(x)$ is the mean pre-bleach intensity (averaged over the scans prior to bleaching). This was further corrected for residual bleaching in the recovery phase as $I(x,t) = I_{norm}(x,t) / I_{norm,ref}(t)$, where $I_{ref}(t)$ is the fluorescence intensity in the reference part of the line that was exempt from the deliberate 10 bleach iterations.

The fluorescence recovery in LineFRAP was described according to[53]

$$I(x,t) = k \sum_{j=0}^{\infty} \frac{(-K_0)^j}{j!} r_{0e} \left[ j r_{0c}^2 + \left(1 + 2j \frac{t}{\tau_r}\right) r_{0e}^2 \right]^{-1/2} + (1-k) I(x,0), \qquad [1]$$

where $K_0$ is the bleaching parameter, $r_{0c}$ is the imaging resolution, and $r_{0e}$ the width of the bleached line. Moreover, the diffusion constant $D$ is obtained from the characteristic recovery time $\tau_r = r_{0e}^2/4D$, $k$ is the mobile fraction and $I(x,0)$ is the fluorescence intensity immediately after bleaching. In our experiments, we set the bleached fraction to be relatively small such that $r_{0e} \approx r_{0c}$ (Supporting Figure S6) which simplifies the equation to

$$I(x,t) = k \sum_{j=0}^{\infty} \frac{(-K_0)^j}{j!} \left[1 + j\left(1 + 2\frac{t}{\tau_r}\right)\right]^{-1/2} + (1-k) I(x,0). \qquad [2a]$$

Implicit to Eq. 2a is

$$I(x,0) = \sum_{j=0}^{\infty} \frac{(-K_0)^j}{j!} [1+j]^{-1/2}, \qquad [2b]$$

which ultimately gives

$$I(x,t) = k \sum_{j=0}^{\infty} \frac{(-K_0)^j}{j!} \left[1 + j\left(1 + \frac{8D}{r_{0e}^2}t\right)\right]^{-1/2} + (1-k)\sum_{j=0}^{\infty} \frac{(-K_0)^j}{j![1+j]^{1/2}}. \quad [3]$$

The underpinning assumptions of this model have been discussed in detail in the original work.[53] Of note here is that the bleaching efficiency can be expected to be homogeneous throughout the entire sample along the optical axis because the gap size between the apposed surfaces is generally much smaller than the confocal depth in the relevant area close to their contact. Moreover, to meet the requirement of fluorescence molecules being uniformly distributed, we averaged over sections along the bleached line that were wider than the extension of the diffusion front $(Dt)^{-1/2}$. Also, we aimed for keeping the bleaching phase sufficiently short to avoid any significant recovery during that phase.

When fitting with Eq. [3] we neglected all terms of $j \geq 6$. This sped up the analysis and had a negligible influence on the results. Moreover, we fixed $r_{0c} = r_{0e} = r_{xy} = 0.50$ µm (see Supporting Figure S6). The three adjustable parameters thus were $D$, $k$ and $K_0$. Normalized $\chi^2$ values typically were around 5 and residuals scattered evenly around 0, indicating a good fit.

## ACKNOWLEDGEMENTS

This work was supported by the European Research Council (Starting Grant #306435 "JELLY" and Proof of Concept Grant #840295 "DIFFUSION" to RPR) and the United Kingdom Biotechnology and Biological Sciences Research Council (BB/R000174/1 to RPR). O. Borisov (University of Pau - CNRS, France), L. Bureau (University Grenoble Alpes - CNRS, France) and P. Torstensson (Chalmers University of Technology, Gothenburg, Sweden) are acknowledged for fruitful discussions on polymer and solid mechanics.

## SUPPORTING INFORMATION AVAILABLE

Supporting Figures S1 to S6, supporting methods and supporting references. This material is available free of charge via the Internet at http://pubs.acs.org.

## AUTHOR CONTRIBUTIONS

RF and RPR conceived the study. All authors contributed to the design of the experiments. RF, DD and RPR invented the PSCM technology. JS produced the proteins. RF, SJ, FB and RPR performed the experiments. All authors contributed to the analysis of the data. RF, DD and RPR wrote the manuscript, and all authors commented on the manuscript.

## COMPETING INTERESTS

European patent application no. 19315169.3 relates to the PSCM technology; the application has been filed on 19/12/2019 with University of Leeds and the Conseil National de la Recherche Scientifique (CNRS, France) as applicants, and RF, DD and RPR as inventors.

**TABLE OF CONTENT FIGURE**

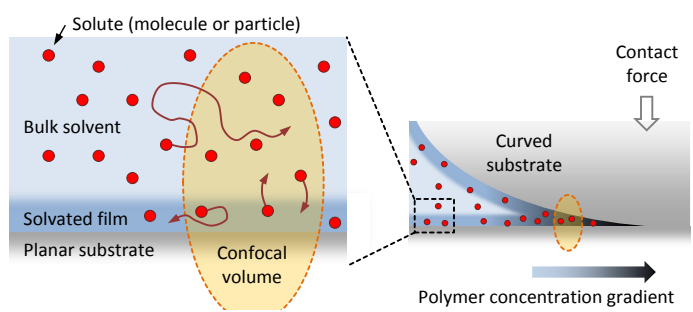

# A method to quantify molecular diffusion within thin solvated polymer films: A case study on films of natively unfolded nucleoporins

Rickard Frost[1], Delphine Debarre[2], Saikat Jana[1], Fouzia Bano[1], Jürgen Schünemann[3], Dirk Görlich[3] and Ralf P. Richter[1,*]

[1]School of Biomedical Sciences, Faculty of Biological Sciences, School of Physics and Astronomy, Faculty of Engineering and Physical Sciences, Astbury Centre of Structural Molecular Biology, and Bragg Centre for Materials Research, University of Leeds, Leeds, LS2 9JT, United Kingdom

[2] Univ. Grenoble Alpes, CNRS, LIPhy, 38000 Grenoble, France

[3]Department of Cellular Logistics, Max Planck Institute for Biophysical Chemistry, 37077 Göttingen, Germany

## SUPPORTING INFORMATION (SI)

# SUPPORTING FIGURES

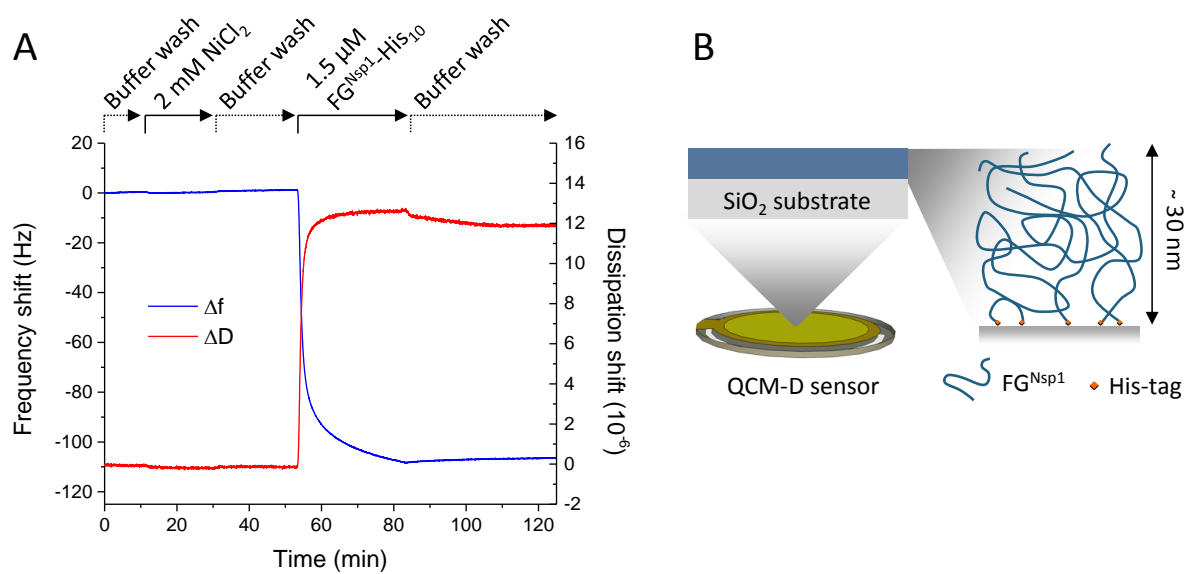

**Figure S1. One-end grafting of Nsp1 FG domains characterized by quartz crystal microbalance (QCM-D).** (**A**) QCM-D data for the binding of $FG^{Nsp1}$-$His_{10}$ to a $Ni^{2+}$-EDTA functionalized surface. To a first approximation, a decrease in the frequency shift relates to addition of mass (including hydrodynamically coupled solvent) to the sensor surface, and the magnitude of dissipation provides information about the softness of the surface adlayer. Solid arrows on top of the graph indicate the start and duration of incubation with sample solution; during remaining times (dashed arrows), the surface was exposed to plain working buffer. Conditions: $NiCl_2$ – 2 mM in working buffer, $FG^{Nsp1}$-$His_{10}$ – 0.16 µM (0.1 mg/mL) in working buffer. The $Ni^{2+}$ ions are too small for the loading of EDTA to be detected. $FG^{Nsp1}$-$His_{10}$ binding is initially rapid and then slows down with a final frequency shift of -108 Hz being reached after 30 min of incubation. The $FG^{Nsp1}$ film is stable to rinsing with working buffer, and the high dissipation indicates it is soft. (**B**) Schematic representation (at three levels of magnification) of a brush of Nsp1 FG domains (right) formed on the silica-coated surface (top left) of a QCM-D sensor (bottom left).

QCM-D experiments were performed using a Q-Sense E4 equipped with Flow Modules (Biolin Scientific, Västra Frölunda, Sweden) under constant flow (20 µL/min) at a working temperature of 22 °C. Data were acquired at several harmonics ($i$ = 3, 5, 7, 9, 11 and 13; corresponding to resonance frequencies of ~15, 25, ... 65 MHz). Normalized frequency shifts, $\Delta f = \Delta f_i/i$, and dissipation shifts, $\Delta D$, for $i$ = 5 are presented only (all other harmonics provided comparable information). Silica coated QCM-D sensors (QSX303) were purchased from Biolin Scientific. The sensors were cleaned (exept for the sonication step), functionalized with EDTA and stored following the same protocol as described in the Methods section for the glass surfaces. Prior to use, the functionalized sensors were rinsed with water and blow dried using $N_2$ gas.

*Estimation of surface coverage and film thickness.* Films of C-terminally grafted Nsp1 FG domains have been characterized extensively in our previous work.[1-3] In particular, Ananth et al{Ananth, 2018 #108} studied the film formation process by QCM-D and by spectroscopic ellipsometry (SE) under comparable mass transport conditions, and from these data we can correlate a QCM-D frequency shift (at $i$ = 5) of -90 ± 10 Hz with a $FG^{Nsp1}$ surface density (measured by SE) of 4.8 ± 0.5 pmol/cm$^2$ (see Fig. S2B in ref. 1). The magnitude of the frequency shift measured here is -108 Hz, implying that a somewhat higher surface density has been attained. On the other hand, we note that $FG^{Nsp1}$ was incubated under flow in the QCM-D assay and that the surface densities on the surfaces used for PSCM measurements are likely to be somewhat reduced because these were incubated in still solution under otherwise identical conditions. Accounting for these uncertainties, we conservatively estimate the $FG^{Nsp1}$ surface density to 5 ± 1 pmol/cm$^2$ for all PSCM measurements. Moreover, a previous extensive analysis by AFM nanoindentation, QCM-D and SE revealed the thickness of films of C-terminally anchored $FG^{Nsp1}$ at around 5 pmol/cm$^2$ to be approximately 30 nm (see Fig. 4B in ref. 3).

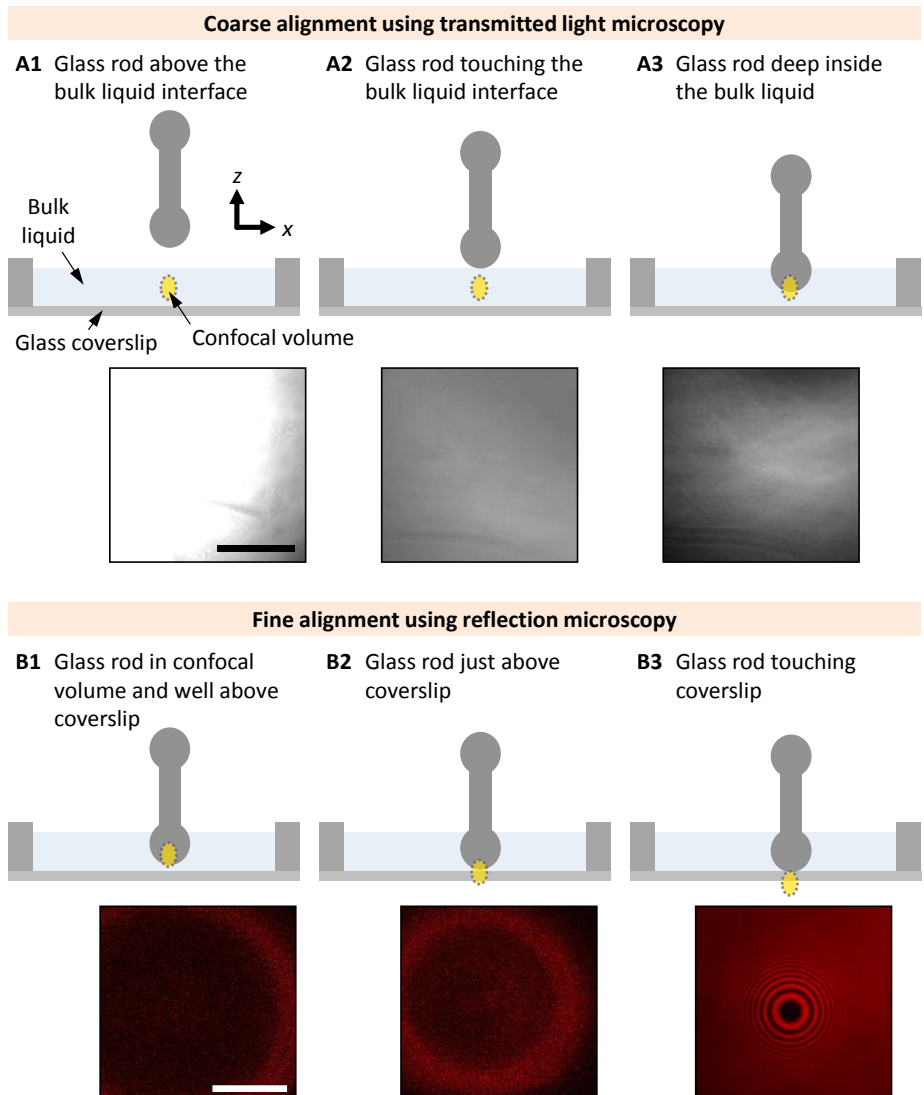

**Figure S2. Aligning the glass rod in the field of view and establishing a sphere-on-plane interface.** The position of the glass cover slip was kept fixed, the position of the glass rod was adjusted in *x*, *y* and *z* by the micromanipulator, and the objective's *z* focus was adjusted as required using the microscope's focus wheel. Prior to coarse approach the position where the glass cover slip interfaced the bulk solution was located: this interface was readily identified by imaging in *reflection* mode (pinhole size of 5 airy units) as a laterally uniform maximum in intensity at this *z* position. **(A1-3) Coarse approach and alignment.** Shown are schematics of the process of lowering the glass rod from air into the bulk liquid along with corresponding *transmission* micrographs of the confocal microscope taken near the rod axis (focus position: 30 μm above the glass cover slip): **(A1)** The hemi-spherical cap of the glass rod was far away from the air-liquid interface. **(A2)** The cap was just in contact with the air-liquid interface. **(A3)** The cap was well immersed in the bulk liquid. Coarse alignment of the glass rod in the centre of the field of view was facilitated by adjusting the detector gain and keeping the bright spot in the centre. **(B1-3) Final approach and fine alignment.** Shown are schematics of the process of approaching the hemi-spherical cap to the planar cover slip along with corresponding *reflection* micrographs near the centre of the cap (pinhole fully open): **(B1)** Initially (focus position: 30 μm above glass cover slip) a faint ring of reflected light indicated the presence of the cap within the confocal volume, and was used to centre the cap in the field of view. **(B2)** As the glass rod and, simultaneously, the focus position were lowered the radius of the ring decreased. **(B3)** Newtonian rings emerged when the hemi-spherical cap was sufficiently close to the glass cover slip (*i.e.* within a distance of a few microns) for interference of light reflected from the two interfaces to occur; interference fringes were best visible with the objective focusing a few microns below the interface, and the micrograph shows the two surfaces in contact. All scale bars are 100 μm; for comparison, the glass rod has a diameter of approximately 6 mm.

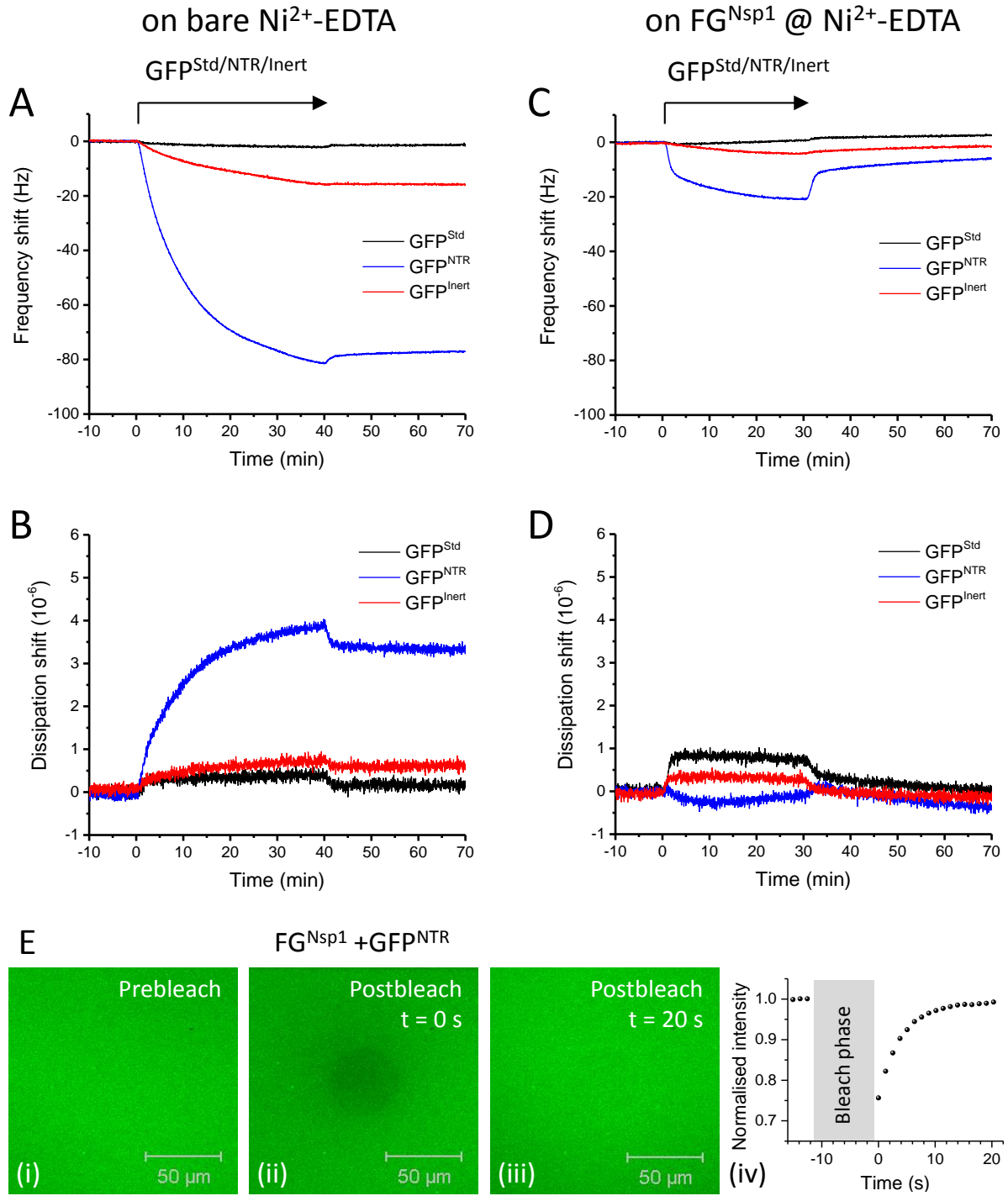

**Figure S3. Interactions of GFP probe molecules with bare and Nsp1 FG domain functionalized surfaces. (A-B)** QCM-D data of the GFP probe molecules (GFP$^{Std}$, GFP$^{NTR}$ and GFP$^{Inert}$, as indicated) when exposed to Ni$^{2+}$-EDTA functionalized silica surfaces. **(C-D)** Data as in (A-B) but for Ni$^{2+}$-EDTA functionalized silica surfaces with an FG$^{Nsp1}$ film (see Figure S1 for the film formation process). Arrows on top of the graph indicate the start and duration of incubation with 2 µM GFP solutions in working buffer; during remaining times, the surface was exposed to plain working buffer. **(E)** Fluorescence micrographs of GFP$^{NTR}$ on a FG$^{Nsp1}$ functionalized surface (corresponding to the scenario in (C-D)). A central circular region (45 µm in diameter) was photo-bleached and the presented images were recorded before bleaching (i), just after bleaching ($t = 0$ s; ii) and 20 s post bleaching (iii); (iv) shows the fluorescence recovery in the bleached area as a function of time. Note that fluorescence signal in this assay stems from GFP$^{NTR}$ located in the FG$^{Nsp1}$ film as well as in the nearby bulk solution, and that recovery in the bulk solution is faster than the time resolution of this disk FRAP assay. Moreover, the rate of recovery of GFP$^{NTR}$ in the film is also enhanced owing to rapid exchange with the bulk solution. This explains the relatively low apparent degree of photo-bleaching at t = 0 min despite the extended bleach phase (marked in grey in (iv)).

From the QCM-D data in (A-D), it is evident that GFP$^{Std}$ did not adsorb to the bare Ni$^{2+}$-EDTA surface or the FG$^{Nsp1}$ film. In contrast, GFP$^{NTR}$ and GFP$^{Inert}$ did adsorb to the bare Ni$^{2+}$-EDTA surface to different extents, and this binding was largely resistant to rinsing in working buffer. However, such undesired binding to the substrate was largely reduced in the presence of the FG$^{Nsp1}$ films: whilst GFP$^{Inert}$ showed only minor binding, GFP$^{NTR}$ binding was pronounced yet largely reversible as expected for a specific interaction with FG$^{Nsp1}$. The FRAP data in (E) confirm that virtually all GFP$^{NTR}$ is mobile and thus reversibly bound in FG$^{Nsp1}$ films.

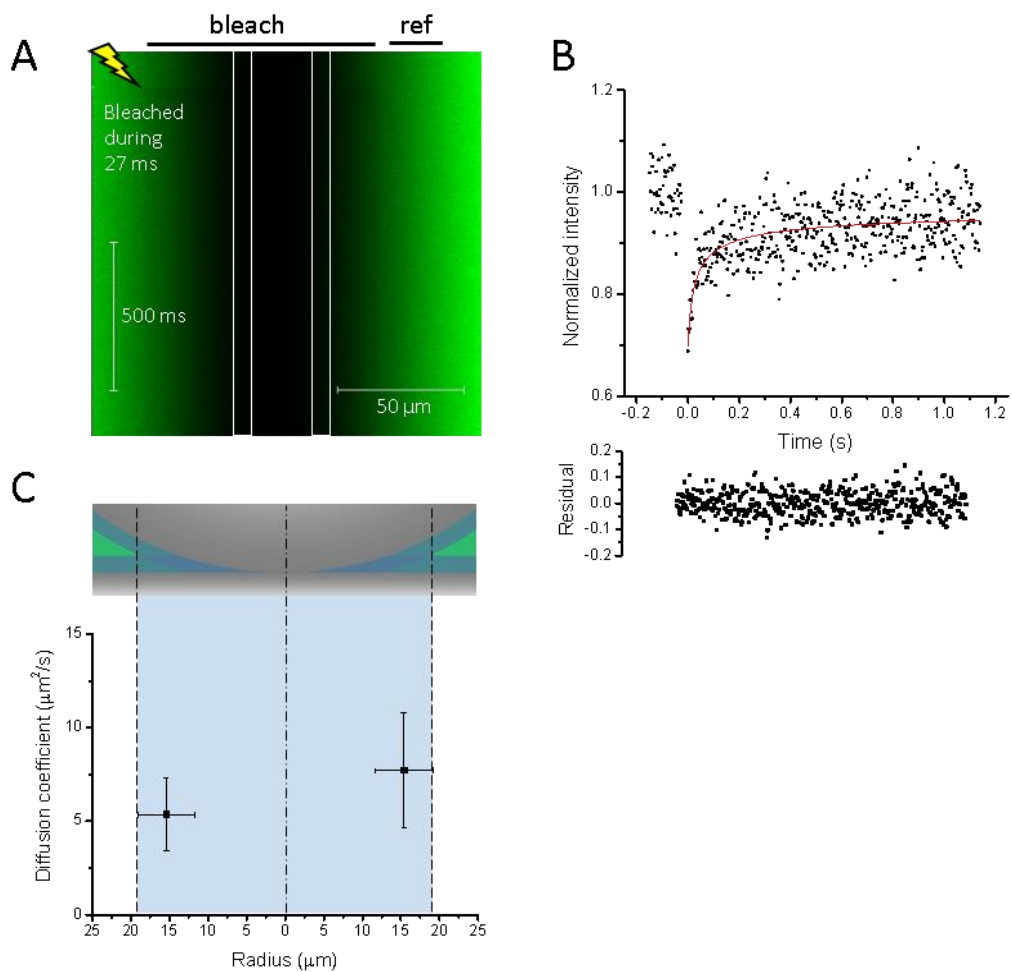

**Figure S4. Diffusion of GFP$^{Std}$ in Nsp1 FG domain films.** Line FRAP data for GFP$^{Std}$ in FG$^{Nsp1}$ films are displayed analogous to Fig. 6A-C. Note that the fluorescence intensity in the region of overlapping FG domain films is reduced for GFP$^{Std}$ as compared to GFP$^{NTR}$, owing to its much lower partition coefficient. To enhance the signal, the present measurement was therefore performed with a 10-fold increased bulk concentration of GFP$^{Std}$ (20 µM) and data were averaged over a wider radial range (7.4 µm; horizontal error bars in C) for line FRAP analysis. The best-fit line in B corresponds to $D = 5.3 \pm 1.9$ µm$^2$/s, $k = 0.91 \pm 0.02$ and $K_0 = 0.52 \pm 0.06$ and the mean of the data in C is $6.5 \pm 1.9$ µm$^2$/s. This value should be considered an estimate because the recovery time ($\tau_r = 11 \pm 4$ ms) was comparable to the bleaching time (27 ms) for this data set.

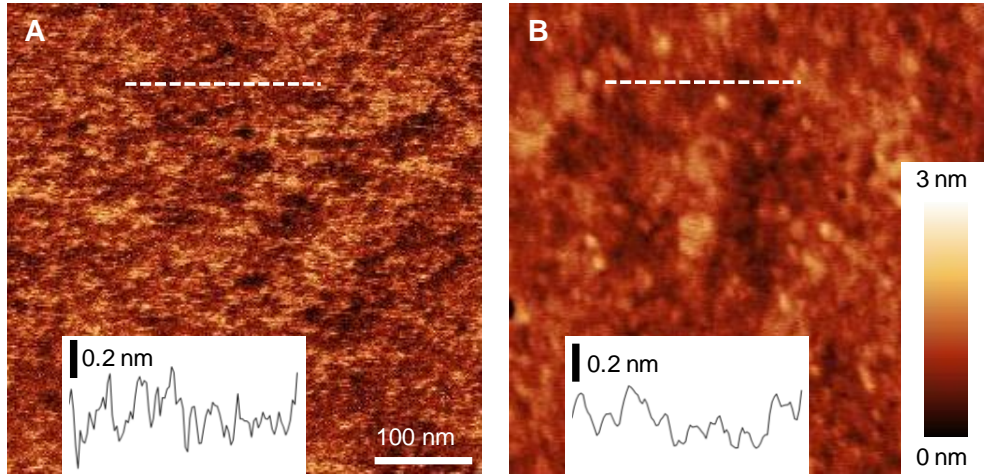

**Figure S5. Glass surfaces are smooth on the nm scale.** Atomic force micrographs (0.5 × 0.5 μm²) of the surface of **(A)** a glass rod (hemi-spherical part), and **(B)** a glass cover slip functionalized with EDTA. Insets show height profiles taken along the white dashed lines. The root-mean-square (rms) roughness values of these surfaces are 0.39 nm and 0.26 nm, respectively, demonstrating the glass surfaces are smooth on the nm scale, without EDTA and after EDTA coating.

Analysis was performed with a Nanoscope Multimode 8 (Bruker, CA, USA) AFM system. Micrographs of the surface topography were acquired in air using Peak Force Tapping mode using sharpened triangular $Si_3N_4$ cantilevers (nominal spring constant 0.06 N/m; NP-S, Bruker). Images were second-order plane fitted (without noise filtering or sharpening) and surface roughness was analysed using Nanoscope Analysis Software.

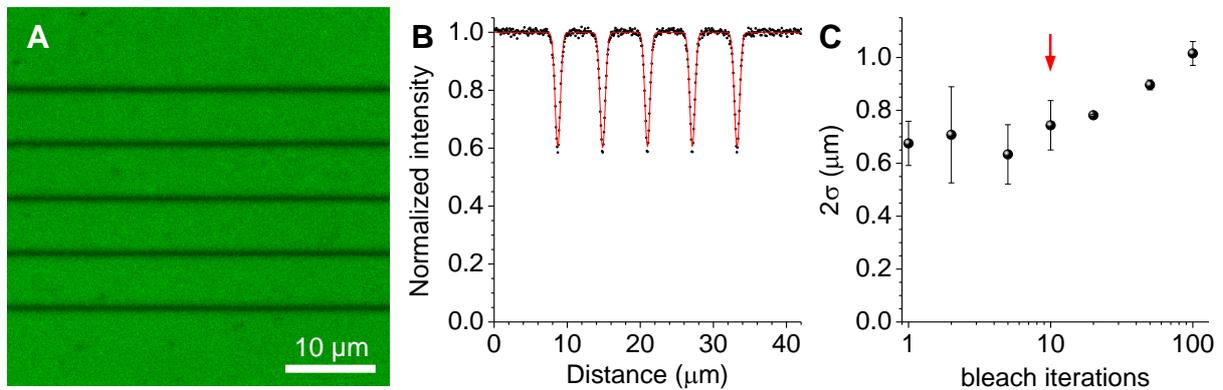

**Figure S6. Determination of the imaging resolution $r_{0c}$ and the width of the bleached line $r_{0e}$.** **(A)** Confocal fluorescence micrograph of a monolayer of GFP with an N-terminal polyhistidine tag ($His_{14}$-GFP)[1] formed on a $Ni^{2+}$-EDTA functionalized surface. This micrograph was obtained after several equidistant lines had been photo-bleached using the same settings as applied during line FRAP. **(B)** Intensity profile obtained by normalizing the data in (A) and subsequent averaging along the horizontal axis (black dots). The red line represents the best fit with a sum of five Gaussian 'holes' and is seen to reproduce the data well. **(C)** Peak width, expressed as $2\sigma$ (where $\sigma^2$ is the variance), obtained from equivalent fits using a range of bleaching iterations (the red arrow highlights the number of iterations used for line FRAP). Error bars represent standard deviations across multiple images ($n = 10$ for 10 bleach iterations, $n = 2$ otherwise). Up to 20 bleach iterations, the peak width does not depend significantly on the number of bleach iterations; this indicates that the bleached line width does not depend on the bleaching level, implying that the bleached line width equals the imaging resolution, over this range. The convolution of the bleached lines with the imaging resolution gives $(2\sigma)^2 = r_{0e}^2 + r_{0c}^2$, and with $r_{0e} = r_{0c}$, we obtain $r_{0e} = r_{0c} = \sqrt{2}\sigma$. Averaging the data in (C) up to 20 bleach iterations gives $2\sigma = 0.71 \pm 0.06\ \mu m$, and thus $r_{0e} = r_{0c} = 0.50 \pm 0.04$ μm.

## SUPPORTING METHODS

**RICM analysis of a hemi-sphere pressing on a planar surface**
*Geometry of the interface.* The geometry of the interface is schematically shown in Fig. S7A. In the absence of force applied by the hemi-sphere, the cover slip is ideally flat ($h_1(r) = 0$) and the end of the rod is ideally hemi-spherical ($h_2(r) = R - \sqrt{R^2 - r^2}$). Because the radius of the hemi-sphere is large ($R = 3$ mm) compared to the size of the area of interest in the image ($r_{max} \approx 100$ µm), the shape of the hemi-sphere can be approximated by a parabola for RICM intensity calculation ($h_2(r) \approx \frac{r^2}{2R}$) and 'nonlocal' curvature effects can be neglected. The phase in the RICM profile can thus be written as

$$\Phi(r) = \frac{2\pi n}{\lambda} \times 2 \times \frac{r^2}{2R}, \qquad [S1]$$

where $n$ is the refractive index of the medium between the surfaces and $\lambda$ is the wavelength of light. Here any defocus is neglected as only very large changes in focus would modify the RICM pattern (*vide infra*).

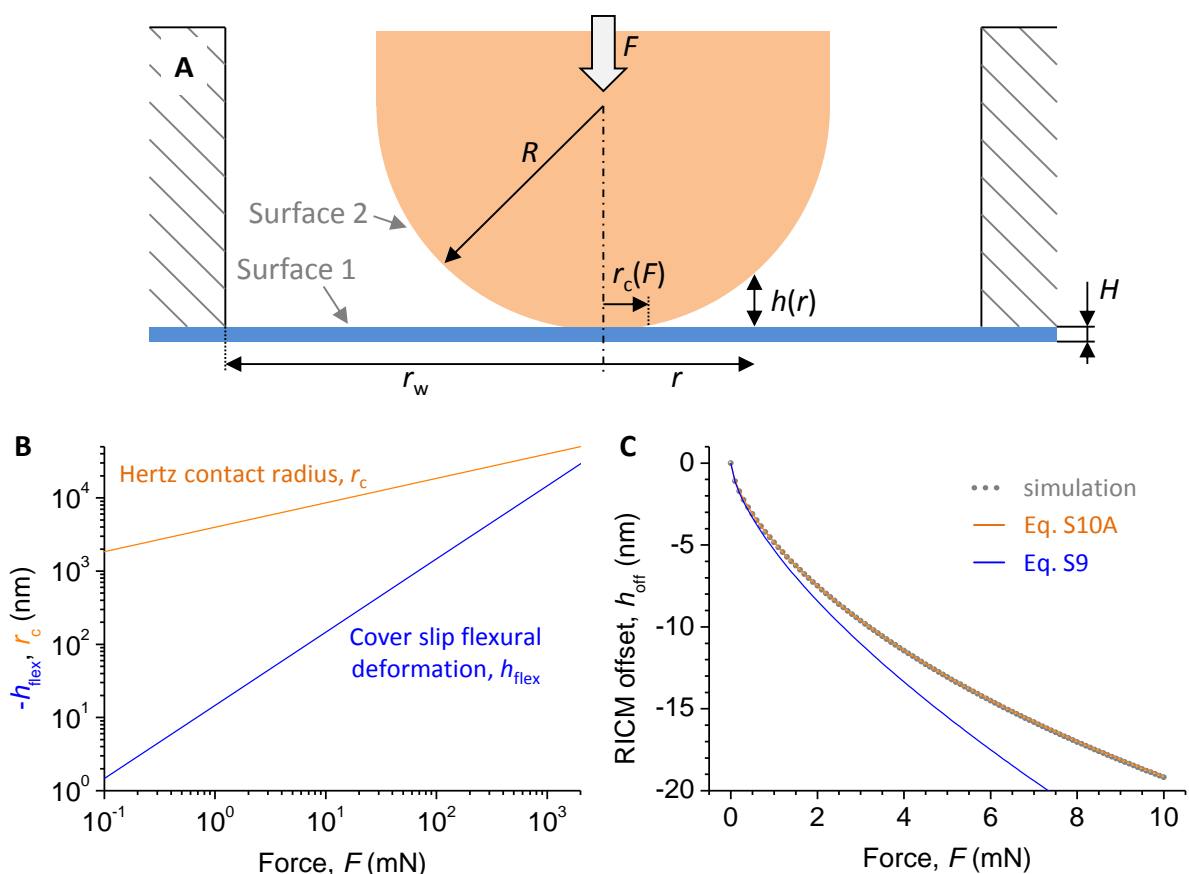

**Figure S7. RICM analysis of a sphere pressing on a planar surface.** (**A**) Schematic of the interface geometry: the planar glass cover slip (#1.5 with thickness $H = 175 \pm 15$ µm) is glued to the sample holder (and hence fixed at the edges) defining a well of radius $r_w = 5$ mm; the hemi-spherical end of a glass rod (radius $R = 3$ mm) presses with a force $F$ at the centre of the cover slip; the area of contact between the two surfaces is a disk of radius $r_c$, and the distance $h$ between the plane and the hemi-sphere increases with the distance $r$ from the centre of the contact area. (**B**) Amplitude of deformation, as a function of applied force, of the cover slip as a whole (blue line) and of the two surfaces around the contact area (orange line). (**C**) Simulated RICM offset *vs.* force applied (grey dots), and fits with Eqs. S9 (blue line) and S10A (orange line).

Applying a force $F$ on the hemi-sphere, two effects will modify the geometry of the setup:

- The cover slip will flex at large scale. The change in height due to flexural deformation is (Eq. 6.6.25 in ref. 4)

$$h_{\text{flex}}(r) = -\frac{F}{16\pi D}\left(r_{\text{w}}^2 - r^2 + 2r^2 \ln\frac{r}{r_{\text{w}}}\right), \qquad [S2]$$

where $D = E_1 H^3/[12(1-\nu_1^2)]$ is the flexural rigidity of the cover slip along with its Young's modulus $E_1$, Poisson ratio $\nu_1$ and thickness $H$. This equation assumes a point-like contact which is not valid around the contact area but gives a good order of magnitude of the large scale deformation. In particular, the maximum deformation is located at the centre of the cover slip and scales as

$$h_{\text{flex}}(0) = -\frac{F r_{\text{w}}^2}{16\pi D}. \qquad [S3]$$

With $E_1 = 72.9$ GPa, $\nu_1 = 0.208$ and $H = 175$ µm (provided by the supplier), we obtain $D = 0.034$ Pa·m³. As shown in Fig. S7B (blue line), this deformation is of small amplitude ($h_{\text{flex}} < 120$ nm) for $F \leq 10$ mN. Moreover, this small vertical deflection is applied over a large distance ($r_{\text{w}} = 5$ mm $\gg 100$ µm, over which the RICM pattern is observed). Hence the contribution of the flexural deformation to the change in the RICM pattern can be neglected.

- At smaller scale around the contact point, both surfaces are deformed and this deformation can be described by Hertz's theory. Fig. S7B shows that this deformation is much more significant, and hence it is the only one considered in the following analyses.

***Hertz contact deformation of the surfaces and resulting RICM pattern.*** A detailed treatment of Hertz contact mechanics can be found, for example, in ref. 5. When applying a force $F$, both the hemi-sphere and the coverslip will deform to give rise to a contact area of radius (from Eqs. 4.22-4.24 in ref. 5)

$$r_{\text{c}}(F) = \left(\frac{3FR}{4E^*}\right)^{1/3} \quad \text{with} \quad \frac{1}{E^*} = \frac{1-\nu_1^2}{E_1} + \frac{1-\nu_2^2}{E_2}, \qquad [S4]$$

where $\nu_j$ and $E_j$ are the Poisson ratios and the Young's moduli, respectively, of the cover slip ($j=1$) and the rod ($j=2$).

Outside the contact area, the deformations are (from Eq. 3.42a in ref. 5)

$$\delta h_1(r) = \frac{E^*}{\pi R}\frac{1-\nu_1^2}{E_1}\left[(2r_{\text{c}}^2 - r^2)\arcsin\frac{r_{\text{c}}}{r} + r_{\text{c}}\sqrt{r^2 - r_{\text{c}}^2}\right], \text{ and} \qquad [S5A]$$

$$\delta h_2(r) = \frac{E^*}{\pi R}\frac{1-\nu_2^2}{E_2}\left[(2r_{\text{c}}^2 - r^2)\arcsin\frac{r_{\text{c}}}{r} + r_{\text{c}}\sqrt{r^2 - r_{\text{c}}^2}\right]. \qquad [S5B]$$

The gap distance is

$$h(r,F) = \begin{cases} 0 & , r \leq r_{\text{c}} \\ \frac{1}{\pi R}\left[(r^2 - 2r_{\text{c}}^2)\left(\frac{\pi}{2} - \arcsin\frac{r_{\text{c}}}{r}\right) + r_{\text{c}}\sqrt{r^2 - r_{\text{c}}^2}\right] & , r > r_{\text{c}} \end{cases} \qquad [S6A]$$

and hence the RICM phase becomes

$$\Phi(r,F) = \begin{cases} 0 & , r \leq r_{\text{c}} \\ \frac{4n}{\lambda R}\left[(r^2 - 2r_{\text{c}}^2)\left(\frac{\pi}{2} - \arcsin\frac{r_{\text{c}}}{r}\right) + r_{\text{c}}\sqrt{r^2 - r_{\text{c}}^2}\right] & , r > r_{\text{c}} \end{cases} \qquad [S6B]$$

From Fig. S7B (orange line) we can estimate that $r_c$ is in the micrometre range for applied forces in the mN range. A good approximation of the RICM pattern (which is typically fitted over the range 0-100 µm) is then obtained by looking at the limit $r \gg r_c$. In this case one obtains

$$\Phi(r, F) = \frac{2\pi n}{\lambda R}(r^2 - 2r_c^2) \quad , r \gg r_c \qquad [S7]$$

Fitting this phase with a quadratic formula including a constant offset $\Phi_0 = \frac{4\pi n}{\lambda} h_{\text{off}}$ (as usually done to account for a distance $h_{\text{off}}$ between the two reflecting surfaces) one thus gets

$$h_{\text{off}} \approx -\frac{r_c^2}{R} \approx -\left(\frac{3F}{4E^*\sqrt{R}}\right)^{2/3} \qquad [S8]$$

Using $R = 3$ mm, $E_1 = 72.9$ GPa, $\nu_1 = 0.208$, $E_2 = 64$ GPa and $\nu_2 = 0.2$ (as provided by the suppliers), one gets $E^* = 35.56$ GPa and

$$h_{\text{off}} \approx -5.3 \text{ nm} \times \left(\frac{F}{\text{mN}}\right)^{2/3}. \qquad [S9]$$

The above analysis is confirmed by a more rigorous fitting of the RICM profile derived from the exact expression of the phase (Eq. S6B). The results are shown in Fig. S7C (grey dots). For small forces ($F < 1$ mN), Eq. S9 fits the simulation well (Fig. S7C, blue line). For larger forces, there is a deviation and the simulation is better fitted by the formula (Fig. S7C, orange line)

$$h_{\text{off}} \approx -5.03 \text{ nm} \times \left(\frac{F}{\text{mN}}\right)^{2/3} \times \left[1 - 0.039\left(\frac{F}{\text{mN}}\right)^{2/3}\right], \qquad [S10A]$$

with an error of less than 0.02 nm across $0 < F < 10$ mN. The simulations also showed that the fit is robust to small errors (a few percent) in $R$ as well as to small differences (a few µm) between the upper surface of the cover slip and the imaging plane (data not shown).

The above analysis was performed for two glass surfaces in direct contact. Equivalent simulations with a 6 nm thick interlayer of $n_{\text{il}} = 1.47$ (*i.e.*, the equivalent of two FG$^{\text{Nsp1}}$ films considering also the EDTA surface functionalisation, *vide infra*) were well fitted by

$$h_{\text{off}} \approx 6 \text{ nm} \times \frac{1.47}{1.334} - 5.06 \text{ nm} \times \left(\frac{F}{\text{mN}}\right)^{2/3} \times \left[1 - 0.030\left(\frac{F}{\text{mN}}\right)^{2/3}\right], \qquad [S10B]$$

with an error of less than 0.25 nm across $0 < F < 10$ mN.

***Determination of the compressive force and interface geometry from RICM.*** Using Eqs. S10 one can readily estimate the compressive force from experimentally measured RICM heights. Eqs. S4 and S5 provide a description of the geometry of the sphere-plane interface for any given applied force.

We note here that the additional FG domain interlayer affects the shape of the $h_{\text{off}}(F)$ curve only marginally. From a comparison of Eq. S10B with Eq. S10A, it is clear that there is a positive constant offset of $6 \text{ nm} \times \frac{1.47}{1.334} \approx 6.6$ nm to $h_{\text{off}}$. Any additional change in $h_{\text{off}}$, however, is marginal: over the relevant force range of 10 mN, it remains below 1 nm. For simplicity, we have neglected the additional change, as it is below the resolution limit of $h_{\text{off}}$ in our setup, but we did take into account the constant offset when calculating the applied force from RICM data.

From the above analysis (Fig. S7C) one can calculate the forces at play for 'soft' contact ($h_{\text{off}} \approx -5$ nm, $F \approx 1$ mN), 'medium' contact ($h_{\text{off}} \approx -10$ nm, $F \approx 3$ mN) and 'hard' contact ($h_{\text{off}} \approx -17$ nm, $F \approx 8$ mN). These values correspond to Hertz contact radii $r_c(F)$ of roughly 4, 6 and 8 µm, respectively (Fig. S7B).

*Generalisation of RICM analysis.* The example above demonstrated how the applied force can be extracted from RICM data for a compressed interlayer of known optical thickness $h_{il}^{opt} = h_{il} n_{il}$ (where $h_{il}$ and $n_{il}$ are the geometrical thickness and the refractive index, respectively, of the optically homogeneous interlayer; for interlayers with a refractive index gradient along the surface normal, the value can be calculated as $h_{il}^{opt} = \int_0^{h_{il}} n_{il}(h) dh$). Analogously, it is also possible to determine the optical thickness of the (compressed) interlayer from the RICM data for a given applied force. Whilst a detailed procedure is not presented here, we highlight that it is generally convenient to make the ansatz

$$h_{off} \approx f(h_{il}, n_{il}) - g(F), \qquad [S11]$$

where $f$ and $g$ are positive functions that describe the effects of the interlayer and of the applied force, respectively. From Eq. S10B, we can identify $f = h_{il}^{opt}/n$ as the 'effective' interlayer thickness. We recall that $g \approx r_c^2/R$ (Eq. S8), *i.e.* to a first approximation $g$ is defined by the radius of contact, as determined by the applied force and the geometry and mechanical properties of the apposed surfaces. $h_{off}$ can be positive or negative, and the sign indicates whether the (positive) effect of the interlayer or the (negative) effect of the force dominate.

**Estimate of compressive forces between hemi-sphere and plane**
As an alternative to the above, we also considered the mechanics of our system to obtain a rough estimate of the applied forces. Whilst the setup as a whole is mechanically complex, we recognised that the lever arm that connects the rod to the micromanipulator is likely to be the most compliant element. We determined the spring constant of the lever arm to be $k_{arm} \approx 250$ N/m using linear regression analysis of its deformation under the load of a set of defined weights (Fig. S8). With this lever spring constant, a lever deflection of 16 μm (*i.e.* from 'soft' to 'hard' contact) corresponds to a difference in force of 4 mN. The order of magnitude compares favourably with the RICM analysis which gave 7 mN over the same range. These values are in reasonable agreement if one considers a resolution limit of ±2 nm in the $h_{off}$ determination.

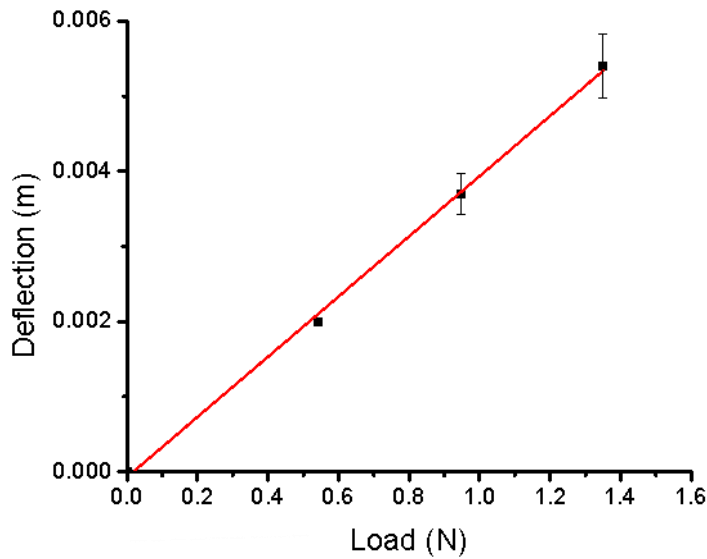

**Figure S8. Spring constant of rod holder.** Deflection *vs.* load of the lever arm holding the glass rod. The loads were applied by attaching weights to the lever arm, and deflections were measured with a ruler. Symbols with errors represent mean and standard deviation of 5 replicas. The red line is the best fit which yields a value of $k_{arm} = 249 \pm 6$ N/m.

**FG domain film thickness under strong compression**

The FG domain of Nsp1 is an intrinsically disordered polypeptide chain. Upon compression, solvent is squeezed out of the FG$^{Nsp1}$ brush, and in the limit of very strong compression it can be expected that the FG$^{Nsp1}$ brush resembles a virtually solvent-free and incompressible polymer melt.[6] With a grafting density $\Gamma = 5 \pm 1$ pmol/cm$^2$, a molecular weight $M_w = 64.1$ kDa of Nsp1 FG domains, and an effective density $\rho = 1.4$ g/cm$^3$ for the compacted polypeptide, we find that the fully compressed film has a thickness of $d_{min} = \Gamma M_w / \rho = 2.3 \pm 0.5$ nm.

To estimate the force required for full compression, we consider the osmotic pressure $\Pi_{osm}$ of a polymer solution according to Flory-Huggins

$$\frac{\Pi_{osm} a^3}{k_B T} = -\ln(1-\varphi) - \varphi - \chi \varphi^2, \qquad [S12]$$

where $k_B T$ is the thermal energy, $a$ the Kuhn segment length, $\chi$ the Flory interaction parameter, and $\varphi$ the polymer volume fraction. The entropy of polymer mixing was here neglected since the polymer chains are confined through grafting. In the limit of strong compression $\varphi$ is approximately constant across the brush, and relates to the brush thickness $d$ as

$$\varphi = \frac{d_{min}}{d}. \qquad [S13]$$

In our experiments a force $F$ is applied and brushes are confined between a hemi-spherical and a planar surface. This can be translated into an equivalent pressure in the geometry of two parallel planar surfaces using Derjaguin's approximation

$$\Pi = \frac{1}{2\pi R} \frac{dF}{dd}, \qquad [S14]$$

where $R$ is the radius of the hemi-sphere. Balancing external and osmotic pressures ($\Pi_{osm} = \Pi$) results in

$$\frac{dF}{dd} = \frac{2\pi R k_B T}{a^3} \left[ -\ln\left(1 - \frac{d_{min}}{d}\right) - \frac{d_{min}}{d} - \chi \left(\frac{d_{min}}{d}\right)^2 \right]. \qquad [S15]$$

Integration of Eq. S4 gives

$$F = \frac{2\pi R k_B T}{a^3} \int_d^\infty \left[ -\ln\left(1 - \frac{d_{min}}{x}\right) - \frac{d_{min}}{x} - \chi \left(\frac{d_{min}}{x}\right)^2 \right] dx$$

$$= \frac{2\pi R k_B T}{a^3} \left[ (d - d_{min}) \ln\left(1 - \frac{d_{min}}{d}\right) + d_{min} - \chi \frac{d_{min}^2}{d} \right]. \qquad [S16]$$

In the limit of $d \to d_{min}$

$$F = \frac{2\pi R k_B T}{a^3} d_{min}(1 - \chi). \qquad [S17]$$

With $a = 0.76$ nm for polypeptide chains, $R = 3$ mm and $\chi > 0$, we find that the brush becomes fully compressed at forces $F > 0.4$ mN. Since PSCM operates at forces in the mN range, we can thus conclude that the brush becomes fully compressed at the centre of the sphere-plane contact area.

We note that the EDTA functionalization of the glass surface that is used to graft the FG$^{Nsp1}$ film makes an additional, minor contribution to the fully compressed film. From the molecular architecture of APTES/EDTA-Ni$^{2+}$, we estimate a thickness between 0.5 and 0.9 nm. Thus, the total thickness of the fully compressed organic film made of two layers of APTES/EDTA-Ni$^{2+}$ and of FG$^{Nsp1}$ (one each per surface) is $6.0 \pm 1.4$ nm.